\documentclass[11pt,onecolumn,twoside]{IEEEtran}

\usepackage{amsmath,amssymb,amsfonts,amsthm,cite,color}

\usepackage[pdftex]{graphicx}

\usepackage{algorithm}
\usepackage{algorithmic}
\usepackage{bm}

\newcommand{\be}{\begin{align}}
\newcommand{\ee}{\end{align}}

\newcommand{\Z}{\mathbb{Z}}
\newcommand{\EX}{\mathbb{E}}
\newcommand{\PR}{\Pr}

\newcommand{\R}{\mathbb{R}}
\newcommand{\vlambda}{\bm{\lambda}}
\newcommand{\C}{\mathcal{C}}
\newcommand{\bu}{\mathbf{u}}

\newtheorem{theorem}{Theorem}

\newtheorem{proposition}[theorem]{Proposition}

\newtheorem{definition}[theorem]{Definition}

\def\mc{\mathcal}
\def\mbf{\mathbf}

\begin{document}
\title{Work Capacity of Freelance Markets: Fundamental Limits and Decentralized Schemes}
\author{Avhishek~Chatterjee, Lav~R.~Varshney, and Sriram Vishwanath%
\thanks{Part of the material in this paper was presented at IEEE INFOCOM, Hong Kong, April-May 2015 \cite{ChatterjeeVV2015}.}
\thanks{A.~Chatterjee is with the Department of Electrical and Computer Engineering, The University of Texas at Austin, Austin, TX 78712 USA. (e-mail: avhishek@utexas.edu).}%
\thanks{L.~R.~Varshney is with the Department of Electrical and Computer Engineering and the Coordinated Science Laboratory, University of Illinois at Urbana-Champaign, Urbana, IL 61801 USA. (e-mail: varshney@illinois.edu).}
\thanks{S.~Vishwanath is with the Department of Electrical and Computer Engineering, The University of Texas at Austin, Austin, TX 78712 USA. (e-mail: sriram@austin.utexas.edu).}%
}
\maketitle

\begin{abstract}
Crowdsourcing of jobs to online freelance markets is rapidly gaining popularity.
Most crowdsourcing platforms are uncontrolled and offer freedom to
customers and freelancers to choose each other. This works well for unskilled jobs 
(e.g., image classification) with no specific quality requirement since
freelancers are functionally identical. For skilled 
jobs (e.g., software development) with specific quality requirements, however, this does not
ensure that the maximum number of job requests is satisfied. In this work
we determine the capacity of freelance markets, in terms of maximum satisfied
job requests, and propose centralized schemes that achieve capacity.
To ensure decentralized operation and freedom of choice for customers 
and freelancers, we propose simple schemes compatible with the operation of 
current crowdsourcing platforms that approximately achieve capacity. 
Further, for settings where the number of job requests exceeds capacity, 
we propose a scheme that is agnostic of that information, but is
optimal and fair in declining jobs without wait. 
\end{abstract}


\section{Introduction}
\label{sec:intro}

Methods and structures for information processing have been changing.  Enabled by the proliferation of modern communication technologies, globalization 
and specialization of workforces has led to the emergence of new decentralized 
models of informational work. Moreover, the millennial generation now entering the workforce 
often favors project-based or job-based work, as in 
crowdsourcing and social production \cite{TapscottW2006,Benkler2006},
rather than long-term commitments \cite{Bollier2011}. Indeed over the last decade, more than 
100 `human clouds' have launched with a variety of structures. These platforms 
serve clients by harnessing external crowds, and global enterprises similarly harness
their internal crowds \cite{BaconBCKPRS2010,VukovicS2012,VarshneyACSOLR2014_arXiv},
making use of human cognitive surplus for information processing \cite{Shirky2010}.

Platforms follow different collective intelligence models 
\cite{MaloneLD2010,BoudreauL2013}, and require different strategies for allocating 
informational work to workers.  In crowdsourcing contest platforms like InnoCentive and TopCoder,
there is \emph{self-selection}: work is issued as an open call and anyone can 
participate in any job; the best submission wins the reward 
\cite{DiPalantinoV2009,ArchakS2009,BoudreauLL2011,RanadeV2012}.
In microtask crowdsourcing platforms like Amazon Mechanical Turk, any worker
is assumed able to do any job and so first-come-first-serve strategies 
are often used; level of reliability may be considered in optimal allocation \cite{KargerOS2014}.
In freelance markets like oDesk and Elance, however, specialized jobs
must be performed by skilled workers: allocation requires careful 
selection from the large pool of variedly-skilled freelancers. 

Freelance market platforms serve as spot markets for labor by matching skills
to tasks, often performing on-demand matching at unprecedented scales.
For example, oDesk had 2.5 million workers and nearly 0.5 million clients
in 2013 \cite{BoudreauL2013}.  Herein we study allocation and scheduling of informational work 
within these kinds of platforms, via a queuing framework.  We aim to establish
fundamental limits through a notion of \emph{work capacity}, and also develop decentralized
algorithms, which are easily-computed, that nearly achieve these performance limits. 

Freelancers may have one or more skills (that are known, cf.~\cite{HoV2012})
and jobs may have multiple parts, called {\em tasks}, that require separate skills.  
Due to job skill requirement variety and limited freelancer ability, it is
often not possible to find a freelancer that meets all requirements for a job: 
a job may have to be divided among freelancers. Moreover, a task in a job may require
so much time that even the task may have to be divided among multiple freelancers. 
There are reputation systems within freelance market platforms,
so freelancers have a reputation level as well as minimum acceptable hourly rate and
skills, which allow worker categorization.
Some freelancers are adaptable in terms of hours available to spend on a particular type of task,
whereas others pre-specify hours available for each kind of task.  Here we consider the non-adaptable
setting where, for example, a freelancer may be available for 20 hours (per week)
of any C++ or Java programming, or may be available for 10 hours (per week) of C++ and 10 hours of Java. 
Studying limits for adaptable freelancers
and designing centralized schemes (and their approximations)
are similar, but the distributed schemes require a different approach.

The objective of the platform is to find a good allocation of jobs (and tasks)
to freelancers. Since working on a task requires synchronization among freelancers, work
can only start when the whole task has been allocated. On the other hand, for some
jobs there are interdependencies between different tasks \cite{OppenheimVC2014} and hence, for these jobs all tasks
must be allocated before the job starts. Moreover, some jobs may require all parts  
to be done by freelancers with the same level of expertise for uniform quality and
money spent.  These considerations lead to concepts of \emph{decomposability} and \emph{flexibility}
that are central to our development.

In the ethos of self-selection, it is desirable for crowd systems to not be centrally controlled, 
but rather for jobs and freelancers to choose each other.  Currently, this may happen randomly or 
greedily. This is clearly not optimal, as the following example illustrates. Consider two types
 of jobs (single task) and two categories of 
freelancers. A type $1$ job can be served by either of the worker categories (example, lower 
reputation requirement) whereas type $2$ jobs can only be served by category $2$ workers. If 
freelancers and jobs are allocated arbitrarily then it may happen that type $1$ uses 
many category $2$ freelancers and many type $2$ jobs remain unserved.

Optimal centralized allocation of informational tasks under the constraints of crowd systems 
is related to hard combinatorial problems such as the knapsack problem.
Compared to scheduling problems in computer science \cite{kleinberg2006algorithm}, 
communication networks \cite{neely2010stochastic,srikantYing2014}, and
operations research \cite{Pinedo2012}, crowd systems face challenges
of freedom of self-selection, need for decentralized operation, and 
uncertainty in resource availability.

Prior works in the information theory, networking, and queueing literatures are similar to our work in terms of theoretical framework, performance metrics, and the nature of performance guarantees, but are not directly related.  The notion of \emph{capacity} of a resource-shared system
where jobs are queued until they are served and the notion of a capacity-achieving resource allocation scheme
for this kind of system came to prominence with the work of Tassiulas and Ephremides \cite{tassiulas1992stability,tassiulas1993dynamic}. The capacity concept and capacity-achieving schemes 
were subsequently developed for applications in communication networks \cite{eryilmaz2005stable,neely2005capacity,neely2010stochastic,srikantYing2014}, 
cloud computing \cite{maguluri2012stochastic}, online advertising \cite{menache2009dynamic,tan2012online}, and
power grids \cite{chen2012scheduling}, among others.  With the advent of cloud services, large-scale systems  have attracted 
significant research interest: resource allocation schemes and their performance
(queueing delays, backlogs, etc.) in the large-scale regimes have been studied 
\cite{tsitsiklis2013queueing,bramson2010randomized,shah2014performance,Wang2014}.

In this paper, our goal is to understand the fundamental limits (capacity) of freelance markets
and ways to achieve this ultimate capacity. We first develop a centralized scheme for achieving these maximum allocations 
where a central controller makes all job allocation decisions.  Given the potential large scale 
of platforms, we also discuss low-complexity approximations of the centralized scheme
that almost achieve the limit.  Finally, with an eye towards giving flexibility to
customers (job requesters) in choosing freelancers, we propose simple decentralized schemes with minimal central computation that
have provable performance guarantees. Further, since job arrival and freelancer availability processes are random 
(and sometimes non-stationary), we also address ways to adapt when the system is operating outside its capacity limits.

\section{System Model}
\label{sec:model}

We first provide formal definitions of the nature of informational work and workers,
and establish notation.

Freelancers (or agents) are of $L$ categories. 
In each category $l \in [L]$, there are $M^{l}$ types of agents
depending on their skill sets and available hours.
There are $S$ skills among agents of all 
categories and types. An agent of category $l$ and type $i$ 
has a skill-hour vector $h^{l}_i$, i.e.\ 
$h^{l}_{i,s}$ available hours for work involving skill $s \in [S]$.

Jobs posted on the platform are of $N$ types. Each type of
job $j \in [N]$ needs a skill-hour service $r_j$, i.e.\ 
$r_{j,s}$ hours of skill $s$. A part of a job of type $j$ involving
skill $s$ is called a $(j,s)$-task if $r_{j,s}>0$, which is the
size of this task.

A job of type $j$ can only be served by
agents of categories $l \in \mc{N}(j) \subset [L]$. This restriction is
captured by a bipartite graph $G=([N], [L], E)$, where a tuple
$(j,l) \notin E \subset [N] \times [L]$ implies that category $l$ agents cannot
serve jobs of type $j$.

On the platform, jobs are allocated at regular time intervals to available
agents, these epochs are denoted by $t \in \{1, 2, \ldots\}$. Jobs that arrive
after epoch $t$ has started are considered for allocation in epoch $t+1$, based on 
agents available at that epoch. Unallocated jobs (due to
insufficient number of skilled agents) are considered again in the next epoch.

Jobs arrive according to a $\mathbb{Z}_+^N$-valued stochastic
process $\mathbf{A}(t)=\left(A_1(t), A_2(t), \ldots, A_N(t)\right)$, where
$A_j(t)$ is the number of jobs of type $j$ that arrive in scheduling 
epoch $t$.

The stochastic process of available agents at epoch $t$ is $\mbf{U}(t)=(\mbf{U}^1(t), \mbf{U}^2(t), \ldots, \mbf{U}^L(t))$. 
For each agent category $l$, $\mathbf{U}^l(t)=\left(U^l_1(t), U^l_2(t), 
\ldots, U^l_{M^l}(t)\right)$ denotes the number of available agents of different types 
at epoch $t$.

We assume processes $\mathbf{A}(t)$ and $\mathbf{U}(t)$ are independent of
each other and that each of these processes is independent and identically distributed 
for each $t$.\footnote{Most of our results can be extended to stationary ergodic processes.} 
We also assume that each of these
processes has a bounded (Frobenius norm) covariance matrix.
Let $\Gamma(\cdot)$ be the distribution of $\mbf{U}(t)$, and let
$\bm{\lambda}=\EX[\mathbf{A}(t)]$ and $\bm{\mu}^l=\EX[\mathbf{U}^l(t)]$
for $l \in [L]$ be the means of the processes.

At any epoch $t$, only an integral allocation of a task (say $(j,s)$) 
is possible. A set of tasks $t_1, t_2, \ldots, t_n$ of size $r_1, r_2, 
\ldots, r_n$ of skill $s$ can be allocated to agents $1, 2, \ldots, m$ only if
available skill-hours for skill $s$ of these agents $h_1, h_2, \ldots, h_m$
satisfy
\[
\sum_{p=1}^n v_{ip} \le h_i, \sum_{q=1}^m v_{qj} \ge r_j, j \in [n], i\in[m]
\]
for some $\{v_{pq}\ge 0\}$.

Whether different tasks of a job can be allocated at different
epochs and across different categories of agents depend on the type 
of the job. 
\begin{definition}
A type of job $j$ is called \emph{non-decomposable (decomposable)} if different tasks comprising it are 
(are not) constrained to be allocated at the same epoch.
\end{definition}
\begin{definition}
A type of job $j$ is called \emph{inflexible (flexible)} if different tasks as well parts
of tasks comprising it are (are not) constrained to be allocated to the same category of agents.
\end{definition}

In a system with only {\em decomposable} jobs, given a set of 
$\{\mathbf{u}^l=(u^l_1, u^l_2, \ldots, u^l_{M^l}):l \in [L]\}$ 
agents (that is, $u^l_i$ agents of category $l$ and of type $i$ within that category), 
a number $a_{j,s}$ of $(j,s)$-tasks can
be allocated {\em only if} there exist non-negative
$\{z^l_{j,s}: (l,j,s) \in [L]\times[N]\times[S]\}$ satisfying
\begin{align}
& \sum_l z^l_{j,s} = a_{j,s}, z^l_{j,s}=0 \ \mbox{if} \ (j,l) \notin E,  \mbox{ for all } j \in [N], s \in [S]\mbox{,} \nonumber \\
& \sum_{j\in[N]} z^l_{j,s} r_{j,s} \le \sum_{i\in[M^l]} u^l_{i} h^{l}_{i,s}  \mbox{, for all } l \in [L], s \in [S]\mbox{.} 
\label{eq:decompCond} 
\end{align}

On the other hand, given a set of $\{\mathbf{u}^l=(u^l_1, u^l_2, \ldots, u^l_{M^l}):l \in [L]\}$ 
agents in a system with only {\em non-decomposable} jobs, $a_j$ jobs of type $j$ (for each $j$) 
can be allocated {\em only if} there exist non-negative $\{z^l_{j,s}: (l,j) \in [L]\times[N]\}$ 
satisfying 
\begin{align}
& \mbox{\eqref{eq:decompCond} and}\ a_{j,s} = a_j \mbox{ for all } j,s\mbox{.} \label{eq:nonDecompCond}  
\end{align}

Intuitively, the conditions imply that required skill-hours 
for the set of jobs is less than the available skill-hours of agents. 
The $\{z^l_{j,s}\}$ capture a possible way of dividing tasks across multiple
category of agents, as they can be interpreted as the number (possibly fraction)
of $(j,s)$-tasks allocated to $l$-category agents.
Note that conditions \eqref{eq:decompCond} and \eqref{eq:nonDecompCond} are necessary for
allocations of decomposable and non-decomposable jobs respectively. These conditions only
imply that there exist possible ways of splitting jobs and tasks across different categories of agents
to ensure integral number of tasks (jobs) are allocated in case of decomposable (non-decomposable) jobs.

For a system with only flexible jobs, different parts of a task can be allocated to different
categories and a category can be allocated parts of tasks. Hence, $\{z^l_{j,s}\}$ can possibly 
take any value in $\R_+^{LNS}$. 
Thus flexible and decomposable (non-decomposable) jobs need to satisfy condition \eqref{eq:decompCond} 
(condition \eqref{eq:nonDecompCond}) which we refer to as \emph{FD (FND)}. 

For inflexible jobs, a necessary condition the allocation must satisfy is 
that each category gets the same integral number of 
$(j,s)$-tasks for all $s, j$, i.e.,
\begin{align}
z^l_{j,s} \in \Z_+ \ \mbox{s.t.} \ z^l_{j,s} = z^l_{j,s'} \mbox{ for all } s, s', j\mbox{.} \label{eq:InflCond}
\end{align}
An allocation of inflexible and decomposable (non-decomposable) jobs needs to satisfy conditions
\eqref{eq:decompCond} (condition \eqref{eq:nonDecompCond}) and \eqref{eq:InflCond}, which we 
refer to as \emph{ID (IND)}. 
 
For simplicity, in this work we focus on
systems with only a single one of these four classes of jobs.\footnote{Extension to combinations of 
multiple classes is not much different but requires more notation.} For brevity we use 
the same abbreviations to refer to class of job, as we use for the necessary conditions. 
Thus, we have FD, FND, ID, and IND systems. 

In crowd systems, the scaling of number of job and agent types, rate of job arrivals, and number of 
available agents is as follows: $\lambda(N) = \sum_{j=1}^N \lambda_j$
scales faster than $N$, i.e.\ $\lambda(N)=\omega(N)$ or $\lim_{N \to \infty} N/\lambda(N)=0$ and
the number of skills $S$ scale slower than $N$, i.e.\ $S=o(N)$. In practice, a job requires
at most a constant number of skills $d$, implying there are $\Omega(S^d)$ possible job types. 
On the other hand, the number of skills of an agent $d'<d$ as a job generally requires more 
diversity than a single agent possesses, implying $M=\sum_l M^l=O(S^{d'})$. $L=O(1)$, as it
relates to variation in reputation levels and hourly rates, and so $M=o(N)$.
Beyond these system scalings seen in practice, we assume $\lambda_j(N)=\omega(1), \forall j 
\in [N]$ and $\sum_{j: r_{j,s}>0} \lambda_j(N)=\Omega\left(N^c\right)$ for all $s \in [S]$, 
for some $c>0$. 
In the sequel, we assume these scaling patterns and refer to them
as {\em crowd-scaling}.

\section{Capacity, Outer Region, and Centralized Allocation}
\label{sec:capacity}
In this section we study the limits of a freelance market with centralized
allocation and present a centralized
algorithm that achieves the limit. We also discuss a simpler upper bound for
the capacity region in terms of first-order statistics of the system. 
These results on ultimate system limits and ways to achieve them
are not only important in their own right, but also serve as benchmarks for 
later discussion of decentralized schemes that provably achieve 
nearly the same limits.

To formally characterize the maximal supportable arrival rate of jobs we introduce some 
more notation.
For each $j \in [N]$, let $Q_j(t)$ be the number of unallocated jobs that are in the crowd system {\em just after} 
allocation epoch $t-1$. As defined above, $A_j(t)$ is the number of jobs of type $j$ that arrive between
starts of epochs $t-1$ and $t$. Let $D_j(t)$ be the number of jobs of type $j$ that have been allocated
to agents at epoch $t$; we call a job allocated only when all parts 
have been allocated. Thus the evolution of the process $Q_j(t)$ can be written as:
\begin{equation}
Q_j(t+1) = Q_j(t) + A_j(t) - D_j(t)\mbox{.} \label{eq:Qevolution}
\end{equation}
Note that at any epoch $t$, at most $Q_j(t)+A_j(t)$ type $j$ jobs can be allocated, as this is the total number of
type $j$ jobs at that time and hence $D_j(t) \le Q_j(t)+A_j(t)$, implying $Q_j(t) \ge 0$. 

\noindent{\bf Notation and Convention.} \emph{We denote the interior and the closure of a set $C$ by $\mathring{C}$ and $\bar{C}$, respectively.
When we say $\vlambda=(\lambda_1, \lambda_2, \ldots, \lambda_N)\in\Lambda\subset\R_+^{NS}$ we mean 
$\vlambda^S=\left(\left(\lambda_1, \lambda_1, \ldots, S \ \mbox{times}\right), 
\left(\lambda_2, \lambda_2, \ldots, S \ \mbox{times}\right), \ldots\right) \in \Lambda$.
Also, whenever we say $\Lambda \subseteq(\supseteq) \Lambda'$ for $\Lambda'\subset \R_+^N$, we mean 
for any $\vlambda\in\R_+^N$, $\vlambda \in \Lambda' \Leftarrow(\Rightarrow)  
\vlambda^S \in \Lambda$.}

\begin{definition}
An arrival rate $\vlambda$ is \emph{stabilizable} if there is a job
allocation policy $\mathcal{P}$ under which 
$\mathbf{Q}(t)=\left(Q_j(t), j \in [N]\right)$ has a finite expectation, 
i.e., $\lim \sup_{t \to \infty} \EX[Q_j(t)] < \infty$, for all $j$.
The crowd system is called {\em stable} under this policy.
\end{definition}
\begin{definition}
$\mathcal{C}_{\Gamma}$, a closed subset of $\R_+^N$ is the \emph{capacity region}
of a crowd system for a given distribution $\Gamma$ 
of the agent-availability process if  any $\vlambda \in \mathring{\C}_{\Gamma}$ 
is stabilizable and any $\vlambda \notin {\C}_{\Gamma}$ is not stabilizable.
\end{definition}

\subsection{Capacity Region and Outer Region}
\label{sec:capOut}
Let us characterize the capacity regions of
different classes of crowd systems. For any given set of available agents
$\bu=\left(u^{l}_{i}: 1 \le i \le M^l, 1\le l\le L\right)$, 
we define the set of different types of tasks ($a_{j,s}$) that can be allocated in a crowd system. 
Note that the necessary conditions to be satisfied for tasks to be allocated are
specific to the class of crowd system.

Using the explicit conditions \eqref{eq:decompCond}, \eqref{eq:nonDecompCond}, and \eqref{eq:InflCond}
for tasks (jobs) to be allocated, we define $C^{{\rm FD}}(\bu)$, $C^{{\rm FND}}(\bu)$, $C^{{\rm ID}}(\bu)$,
and $C^{{\rm IND}}(\bu)$ as the set of tasks that can be allocated in FD, FND, ID, and IND
systems respectively for given availability $\bu$. We denote these sets
generically by $C(\bu)$ and refer to conditions FD, IFD, FND, and IND generically  
as {\em crowd allocation constraint} or \emph{CAC}.
\[
C(\bu) := \left\{\left(a_{j,s} \in \Z_+\right): 
\exists \left(z^l_{j,s}\right) \ \mbox{satisfying CAC}\right\},
\]
and $\C(\bu)$ is the convex hull of $C(\bu)$.

The following
theorem generically characterizes capacity regions of different crowd systems.

\begin{theorem}
\label{thm:capacityGen}
Given a distribution $\Gamma$ of agent-availability, i.e., $\Gamma(\bu)=\PR\left(\mbf{U}(t)=\bu\right)$, 
for a $\vlambda \notin \bar{\C}({\Gamma})$ there exists no policy under which the crowd system 
is stable, where
\[
\C({\Gamma})=\left\{\vlambda=\sum_{\bu \in \Z_+^{M}} \Gamma(\bu) \vlambda(\bu) : M=\sum_l M^l, \vlambda(\bu) \in \C(\bu)\right\}\mbox{.}
\]
For FD, FND, and IND systems, for any $\vlambda \in \mathring{\C}({\Gamma})$ there exists a 
policy such that the crowd system is stable. 
\end{theorem}
\begin{IEEEproof}
See Appendix~II.\ref{sec:thm:capacityGen}.
\end{IEEEproof}
This implies that for FD, FND, and IND systems capacity region $\C_{\Gamma}=\bar{\C}({\Gamma})$ and 
for ID systems $\C_{\Gamma} \subseteq \bar{\C}(\Gamma)$ (possibly strict). 
Note that the 
conditions FD, FND, ID, and IND (generically CAC) are necessary conditions for a valid allocation.
The above theorem implies these conditions are also sufficient, except for ID systems.
In Sec.~\ref{sec:multiCategory}, we present an alternate characterization of the capacity 
region for inflexible systems.

Note $\C({\Gamma})$ depends on the distribution of agent availability $\Gamma$, but it is hard
to obtain this distribution for large and quickly-evolving systems in practice.
Hence, a characterization in terms of simpler system statistics is of use. Below is a characterization of a region beyond which no
arrival rate can be stabilized. Borrowing terminology from multiterminal Shannon theory, 
we call this the \emph{outer region}.

For any set $J \subset [N]$, define
$\mc{N}(J)=\{l \in [L]: \ \exists j \in J \ \mbox{s.t.} \ (j,l) \in E\}$ and the closed subset
of $\R_+^N$,
\[
\C^{out}_{\bm{\mu}} = \left\{\vlambda: \forall J \subset [N], \forall s, 
             \sum_{j \in J} \lambda_j r_{j,s} \le \sum_{l \in \mc{N}(J)} \sum_{i \in M^l} \mu^{l}_{i} h^{l}_{i,s}\right\}.
\]

\begin{theorem}
\label{thm:outerGen}
For any distribution $\Gamma$ with mean $\bm{\mu}$, $\C_{\Gamma} \subseteq \C^{out}_{\bm{\mu}}$.
\end{theorem}
\begin{IEEEproof}
See Appendix~II.\ref{sec:thm:outerGen}.
\end{IEEEproof}
In general, $\C_{\Gamma}$ is a strict subset of $\C^{out}_{\bm{\mu}}$ because $\C^{out}_{\bm{\mu}}$ only captures the balance of skill-hours in the crowd-system,
i.e. average skill-hours requirement is no more than average availability, but 
partial allocation of a task is not acceptable
in a crowd system. Moreover, for non-decomposable jobs
all tasks of a job have to be allocated simultaneously. Hence, 
meeting an average skill-hour balance criterion may be far from being 
sufficient for stability. For inflexible systems the requirements are even
stricter, which is likely to increase the gap between the outer region 
and the true capacity region. In Sec.~\ref{sec:multiCategory} we present a tighter 
outer region for inflexible systems.

In certain scenarios $\C^{out}_{\bm{\mu}}$ may be non-empty when $\C_{\Gamma}$ is empty. For example, 
consider a simple non-decomposable crowd system with $N=L=1$ and $M^1=S=2$. 
Let each job require $1$ hour of both skills,
type $i$ agents have only $1$ hour available for skill $i$ and none for other skills, 
$\mbf{U}^1(t)$ be uniformly distributed on $\{(0,10),(10,0)\}$, and $\vlambda=(4,4)$.
Then clearly $\vlambda \in \C^{out}_{\bm{\mu}}$, but note that at any time there is only one type of
skill available, hence no job can be allocated. This implies $\C_{\Gamma}=\emptyset$.

\subsection{Centralized Allocation}
\label{sec:allocation}
Though there exists a policy for each $\vlambda \in \mathring{\C}_{\Gamma}$ that stabilizes the system, 
these policies may
differ based on $\vlambda$ and may depend on the job-arrival and agent availability statistics. 
Changing policies based on arrival rate and statistics is 
not desirable in practical crowd systems due to the significant overhead. 
Below we describe a centralized statistics-agnostic allocation policy which stabilizes any 
$\vlambda \in \mathring{\C}_{\Gamma}$. Later we discuss computational cost of this policy for different 
classes of crowdsourcing system and present simpler distributed (or almost distributed) 
schemes with provable performance guarantees under some mild assumptions.

To describe the scheme we introduce some more notation. Let $Q_{j,s}(t)$ be the number of $s$-tasks (skill $s$)
of type $j$ jobs just after the allocation epoch $t-1$ and let $D_{j,s}(t)$ be the number of $s$-tasks (skill $s$)
of type $j$ jobs allocated at epoch $t$. Then
\[
Q_{j,s}(t+1) = Q_{j,s}(t) + A_j(t) \ 1\left(r_{j,s}>0\right) - D_{j,s}(t) \mbox{.}
\]

Note that for all $t$, due to the CAC condition on allocation, $D_{j,s}(t) \in C(\mbf{U}(t))$. 
Moreover, there is an additional restriction that 
$D_{j,s}(t) \le  Q_{j,s}(t) + A_j(t)$, as there are $Q_{j,s}(t) + A_j(t)$ part $s$ of job type $j$ 
in the system at that time, which in turn implies $Q_{j,s}(t) \in \Z_+$ for all $t$. 
Note that as $D_{j,s}(t) \in C(\mbf{U}(t))$, for non-decomposable systems 
$Q_{j,s}=Q_{j,s'}$ for all $j, s, s'$, whereas for decomposable systems they may
differ.

\begin{algorithm}[h]
\caption{MaxWeight Task Allocation (MWTA)} \label{alg:SRMW}
\begin{algorithmic}
\STATE Input: $\{Q_{j,s}(t):j \in [N],s \in [S]\}$, $\mathbf{A}(t)$ and $\mathbf{U}(t)$ at $t$
\STATE Output: Allocation of jobs to agents
\STATE {\bf MaxWeight}
\begin{align}
& \left(\hat{z}^l_{j,s}(t): l, j, s\right) = \arg \max_{\left(z^l_{j,s}\in \Z_+: l, j, s\right)} \sum_{j,s}   Q_{j,s}(t) \Delta_{j,s} \nonumber \\
& \mbox{s.t.}  \left(z^l_{j,s}\right) \ \mbox{satisfy CAC with} \                                a_{j,s} = \Delta_{j,s}(t) \forall j,s. \nonumber
\end{align} 

\STATE {\bf Task Allocation}
\FOR{$j=1:N$}
\STATE Order $j$-type jobs arbitrarily, $O_j$ 
\FOR{$s=1:S$}
\STATE Use order $O_j$ among non-zero $(j,s)$-tasks
\STATE $l=1$
\WHILE{$l\le L$ and $\sum_{k=1}^{l-1} z^k_{j,s}< Q_{j,s}(t) + A_j(t)$}
\STATE Allocate $[\sum_{k=1}^{l-1}z^{k}_{j,s}:\sum_{k=1}^l z^{l}_k]$ $(j,s)$ tasks to category $l$.
Here tasks $[x:x+y]$  are task set $I=\{\lceil x \rceil, \cdots \lfloor x+y \rfloor\}$ 
(in the ordering $O_j$), $(\lceil x \rceil - x)$ fraction of task $\lceil x 
\rceil$ and $1 + x + y - \lceil x+y \rceil$ fraction of
task $\lceil x+y \rceil$.
\STATE $l\leftarrow l+1$
\ENDWHILE
\ENDFOR
\ENDFOR
\FOR{$l=1:L$}
\STATE Order agents of category $l$ arbitrarily
\FOR{$s=1:S$}
\STATE Agents pick maximum (as per availability constraint) 
tasks (or part) in order from $\sum_j \min(z^l_{j,s},Q_{j,s}(t) + A_j(t)) r_{j,s}$
hours
\ENDFOR
\ENDFOR
\end{algorithmic}
\end{algorithm}

We propose the MaxWeight Task Allocation (MWTA) policy, Alg.~1, to allocate tasks to agents at epoch 
$t$ based only on the knowledge of
$\mbf{Q}(t)$, $\mbf{A}(t)$, and $\mbf{U}(t)$, and therefore statistics-agnostic.  It is based on MaxWeight matching \cite{neely2010stochastic,srikantYing2014}.

It is apparent that the MaxWeight part of the algorithm
finds a $\{z^l_{j,s}\}$ that satisfies CAC. The following theorem implies 
that MWTA allocates tasks optimally.
The proof of the theorem is based on adapting the proof of optimality of the
MaxWeight algorithm under the constraints and assumptions of crowd 
systems. It implicitly relies on the following result.
\begin{proposition}
\label{prop:TA}
For any $\bu$ and $\mbf{Q}$, and $\{z^l_{j,s}\}$ satisfying CAC, the Task Allocation part of 
MWTA (Alg.~1) gives a feasible allocation for FD, FND, and IND systems.
\end{proposition}
\begin{IEEEproof}
See Appendix II.\ref{sec:prop:TA}.
\end{IEEEproof}
\begin{theorem}
\label{thm:SRMW}
MWTA (Alg.~1) stabilizes FD, FND, or IND crowd systems for any arrival rate 
$\vlambda \in \mathring{\C}_{\Gamma}$ (for respective $\C_{\Gamma}$).
\end{theorem}
\begin{IEEEproof}
See Appendix II.\ref{sec:thm:SRMW}.
\end{IEEEproof}

\section{Single-category Systems and Decentralized Allocations}
\label{sec:SCSyst}
There is effectively a single category of agents in many platforms with a large population of
new freelancers, whose reputations are based on evaluation tests for skills
and who are paid at a fixed rate.
Hence designing efficient allocation schemes for single-category systems 
are of particular interest, as this population of agents are significant in
ever-evolving crowd systems. Insights drawn from single-category systems
are also useful in controlling multi-category systems, Sec.~\ref{sec:multiCategory}.

For a single category system ($L=1$), note that $z^1_{j,s}=a_{j,s} \in \Z_+$ and hence the
feasibility condition \eqref{eq:decompCond} reduces to:
\[
\sum_{j} a_{j,s} r_{j,s} \le \sum_i u_{i} h_{i,s} \mbox{ for all } s\in[S], a_{j,s}\in \Z_+  \mbox{,}
\]
with condition \eqref{eq:nonDecompCond} additionally requiring $a_{j,s}=a_{j,s'}$ for all
$j, s, s'$. Thus, $C(\bu)$ is the set of $\{a_{j,s}\}$ satisfying the above conditions
for respective classes of jobs and $\C(\Gamma)$ 
is the weighted (by $\Gamma(\bu)$) sum of convex hulls of $C(\bu)$s, 
here $\C_{\Gamma}=\bar{\C}(\Gamma)$.

$\C^{out}_{\bm{\mu}}$ has a simple characterization as well. 
As for any $j \in [N]$, $(j,1) \in E$, and $\mc{N}(J)=1$ for all $J \subset [N]$,
$\sum_{l \in \mc{N}(J)} \sum_{i \in [M^l]} \mu^{l}_i h^{l}_{i,s} = \sum_{i \in [M]} \mu_i h_{i,s}$.
Thus it is sufficient to satisfy the inequality for $J=[N]$, and hence,
$\C^{out}_{\bm{\mu}} = \left\{ \vlambda: \sum_{j \in [N]} \lambda_j r_j \le  \sum_{i \in [M]} 
\mu_i h_{i}\right\}$.

The MaxWeight computation in MWTA for single-category systems turns out to be the 
following integer linear program (ILP), which is related to knapsack problems.
\begin{align}
& \arg\max_{\{\Delta_{j,s}:j,s\}} \sum_{j,s}   Q_{j,s} \Delta_{j,s} \\ 
& \mbox{s.t.} \ \sum_{j} \Delta_{j,s} r_{j,s} \le \sum_i u_{i} h_{i,s} \forall s\in[S], \nonumber \\
% \Delta_{j,s}=a_{j,s}\forall j, s \notag \\
& \qquad \Delta_{j,s} = \Delta_{j,s'}, \forall s, s', j \ (\mbox{only for ND})  \notag
\label{eq:MWforSC} 
\end{align}

For decomposable and non-decomposable systems, this is a
single knapsack and multi-dimensional knapsack problem \cite{kellerer2004knapsack}, respectively,
and hence NP-hard. There exist fully polynomial time approximations (FPTAS) for
single knapsack, whereas for multi-dimensional knapsack only polynomial time approximations
(PTAS) are possible \cite{kellerer2004knapsack}. With this approximation, say $1-\epsilon$, the MWTA policy
stabilizes $(1-\epsilon)\mathring{\C}_{\Gamma}=\{\vlambda: \frac{\vlambda}{1-\epsilon} \in 
\mathring{\C}_{\Gamma}\}$. Also,
note that for large crowd systems each $\lambda_i$ is large and hence stabilizing
any $\vlambda$ with $\vlambda+\mbf{1} \in \mathring{\C}_{\Gamma}$ is almost optimal. The above ILP can
be relaxed to obtain a linear program, an allocation based on which achieves this approximation (see
Appendix I.A for details).

\subsection{Decentralized Allocations}
Now we show that due to the structure of the crowd allocation problem and the fact
that crowd systems are large, simple allocation schemes with minimal centralized control
achieve good performance under mild assumptions on arrival and availability processes.
Interestingly, though the centralized optimal allocation requires solving a knapsack
problem at each epoch and greedy schemes are known to be sub-optimal for knapsack
problems \cite{kellerer2004knapsack}, we propose two simple greedy schemes that are almost 
optimal with good performance guarantees. One of them, called GreedyAgent allocation 
provably performs well for decomposable systems and offers the freedom of selection
to freelancers. Another, called GreedyJob allocation
has provable performance guarantees for both decomposable and non-decomposable systems
while allowing customers (job requesters) the freedom of selection.
Thus, in some sense, this shows that though greedy algorithms can be suboptimal for 
an arbitrary allocation instant (at each epoch), for a dynamical system over long time,
its performance is good. 
\begin{algorithm}
\label{alg:greedyAgent}
\caption{GreedyAgent Allocation}
\begin{algorithmic}
\STATE Input: $\mbf{A}(t)$
\STATE Output: Job to agent allocations
\STATE $\mc{A}$: set of agents, $\mc{T}$: set of tasks
\WHILE{$\mc{A}$ and $\mc{T}$ non-empty}
\STATE Agents in $\mc{A}$ contend (pick random numbers) and $a$ wins
\FOR{each skill with non-zero skill hour}
\STATE $a$ picks as many integral tasks as it can pick
\IF{$a$ has remaining available hour}
\STATE $a$ Picks from remaining parts of the partially allocated task
\IF{$a$ has remaining available hour}
\STATE $a$ picks part of any unallocated task
\ENDIF
\ENDIF
\STATE Remove fully allocated tasks from $\mc{T}$
\ENDFOR
\STATE $\mc{A}=\mc{A} \backslash \{a\}$
\ENDWHILE
\STATE Tasks with partial allocations are not allocated
\end{algorithmic}
\end{algorithm}

In GreedyAgent allocation (Alg. 2), agents themselves figure out
the allocation via contention, without any central control. Agents need no knowledge about the agent population, but do need information 
on the available pool of jobs and have to agree on certain norms. 
In most freelance market platforms, this information is
readily available, and so an algorithm like this is natural. As expected, this
scheme may not be able to stabilize any arrival rate in $\mathring{\C}_{\Gamma}$ for any ergodic
job-arrival and agent-availability processes, but it has good 
theoretical guarantees under some mild assumptions on the job arrival and agent 
availability processes.

\begin{definition}
A random variable $X$ is
\emph{Gaussian-dominated} if $\EX[X^2] \le \EX[X]^2+\EX[X]$ and 
for all $\theta \in \R$, $\EX[e^{\theta \left(X-\EX[X]\right)}] \le 
\exp\left\{\frac{(\left(\EX[X^2]-\EX[X]^2\right) \theta^2}{2}\right\}$
\end{definition}
\begin{definition}
A random variable $X$ is \emph{Poisson-dominated} if
for all $\theta \in \R$, $\EX[e^{\theta \left(X-\EX[X]\right)}] \le e^{\EX[X] (e^\theta-\theta-1)}$.
\end{definition}

Note that these domination definitions imply that the variation of the random variable around its mean is
dominated in a moment generating function sense by that of a Gaussian or Poisson 
random variable.\footnote{Domination in this sense is used in bandit problems 
\cite{bubeck2012regret}.} Such a property is satisfied by many distributions including
Poisson and binomial that are used to model arrival processes for many systems,
e.g., telephone networks, internet, call centers, and some freelance markets 
\cite{BaconBCKPRS2010,VukovicS2012}. It is not hard to show that sub-Gaussian distributions 
(standard in machine learning \cite{keshavan2009matrix}) that 
are symmetric around their mean, are Gaussian-dominated.

The following theorem gives a guarantee on the performance of
GreedyAgent, under mild restrictions on the job-arrival and agent-availability
processes.  Independence assumptions are not too restrictive for
large crowd systems, where jobs and agents may come from different well-separated geographies
or organizational structures. 
\begin{theorem}
\label{thm:greedyAgent}
If the arrival processes $\{A_j(t)\}$ and the agent availability processes
$\{U_i(t)\}$ are i.i.d.\ across time and independent across types (jobs and agents) and 
all these processes are Gaussian-dominated (and/or Poisson-dominated), then
for any given $\alpha\in(0,1]$, there exists an $N_{\alpha}$ such that GreedyAgent allocation 
stabilizes any arrival rate $\vlambda \in (1-\alpha) \C^{out}_{\bm{\mu}} := 
\left\{\vlambda: \frac{1}{1-\alpha} \vlambda \in \C^{out}_{\bm{\mu}} \right\}$ 
for any single-category decomposable crowd system with $N \ge N_{\alpha}$. Moreover, for any arrival rate 
in $(1-\alpha) \C^{out}_{\bm{\mu}}$, at steady state, after an allocation epoch, 
the number of unallocated tasks is $O(S \log N)$ with 
probability $1-o\left(\frac{1}{N^2}\right)$.
\end{theorem}
\begin{IEEEproof}
See Appendix II.\ref{sec:thm:greedyAgent}.
\end{IEEEproof}
As $\C_{\Gamma} \subseteq \C^{out}_{\bm{\mu}}$, this implies that the greedy scheme
stabilizes an arbitrarily large fraction of the capacity region, under the assumptions
on the arrival and availability processes. As $S=o(N)$, more specifically
$O(N^c)$ for $c<1$, the above bound on number of jobs imply that there are $o(N)$
unallocated tasks at any time. This in turn implies that unallocated tasks per
type (average across types) is $o(1)$, i.e., vanishingly small number of tasks per type
are unallocated as the system scales.

In GreedyAgent, there is no coordination among agents while picking tasks
within jobs. Hence in a non-decomposable system, many tasks may be picked by agents 
but only few complete jobs are allocated. As more and more jobs accumulate,
the chance of this happening increases, resulting in more accumulation. This
can result in the number of accumulated jobs growing without bound, as formalized below.
\begin{proposition}
\label{prop:greedyAgentNDunstable}
There exists a class of non-decomposable crowd systems with Poisson-dominated 
(as well as Gaussian-dominated) distributions of arrival and availability, 
such that the system is not stable under GreedyAgent allocation.
\end{proposition}
\begin{IEEEproof}
See Appendix II.\ref{sec:prop:greedyAgentNDunstable}.
\end{IEEEproof}

Hence, we propose another simple greedy scheme that works for both decomposable and
non-decomposable systems. The GreedyJob allocation
scheme (Alg.~3) is completely distributed and hence 
a good fit for crowd systems. GreedyJob has similar performance guarantees for both decomposable and
non-decomposable systems as GreedyAgent has for decomposable systems only.

\begin{algorithm}
\label{alg:greedyJob}
\caption{GreedyJob Allocation}
\begin{algorithmic}
\STATE Input: $\mbf{U}(t)$
\STATE Output: Job to agent allocations
\STATE $\mc{J}$: set of all jobs
\WHILE{Available skill-hours of agents and $\mc{J}\neq \emptyset$}
\STATE Jobs in $\mc{J}$ contend (pick random numbers) and $J$ wins
\IF{$J$ finds agents to allocate all tasks}
\STATE Allocate to those agents
\ELSE
\STATE $J$ does not allocate anything
\ENDIF
\STATE $\mc{J}=\mc{J}\backslash \{J\}$
\ENDWHILE
\end{algorithmic}
\end{algorithm}

\begin{theorem}
\label{thm:greedyJob}
If the arrival processes $\{A_j(t)\}$ and the agent availability processes
$\{U_i(t)\}$ are i.i.d.\ across time and independent across types (jobs and agents), 
all these processes are Gaussian-dominated (and/or Poisson-dominated)
and $\forall s,s', |\sum_i \mu_i h_{i,s} - \sum_i \mu_i h_{i,s'}|$ is $O(\mbox{subpoly}(N))$, 
then for any given $\alpha\in(0,1]$, $\exists N_{\alpha}$ such that GreedyJob allocation 
stabilizes any arrival rate $\vlambda \in (1-\alpha) \C^{out}_{\bm{\mu}} := 
\left\{\vlambda: \frac{1}{1-\alpha} \vlambda \in \C^{out}_{\bm{\mu}} \right\}$ 
for any single-category crowd-system with $N \ge N_{\alpha}$. Moreover, for any arrival rate 
in $(1-\alpha) \C^{out}_{\bm{\mu}}$, at steady state, after an allocation epoch, 
total number of unallocated jobs (adding all types) is  $O(\log N)$ with 
probability $1-o\left(\frac{1}{N^2}\right)$.
\end{theorem}
\begin{IEEEproof}
See Appendix II.\ref{sec:thm:greedyJob}.
\end{IEEEproof}
In Sec.~\ref{sec:multiCategory} we propose a decentralized scheme for multi-category systems 
that uses the two single-category decentralized schemes as building blocks.
At the end of Sec.~\ref{sec:multiCategory} we briefly 
discuss the suitability of these decentralized schemes for 
crowd systems in terms of implementability on crowd platforms.

\section{Multi-Category Systems}
\label{sec:multiCategory}
Sec.~\ref{sec:capacity} characterized the capacity region and developed an optimal centralized scheme for
crowd systems in, whereas Sec.~\ref{sec:SCSyst} discussed simple decentralized schemes 
for single-category systems. Here we return to multi-category systems, 
briefly discussing computational aspects of MWTA, followed by an alternate approach to
the capacity and outer region of inflexible systems that yields a simple optimal 
scheme. We also present a decentralized scheme based on insights from the
optimal scheme and the decentralized allocations in Sec.~\ref{sec:SCSyst}.

The MWTA scheme, which is throughput optimal for FD, FND, and IND systems,
involves solving an NP-hard problem for multi-category systems.
For multi-category systems this is from the
general class of packing integer programs, for which constant factor
approximation algorithms exist under different assumptions on the 
problem parameters \cite{chekuri2004multidimensional}. 
These assumptions do not generally hold 
for MaxWeight allocation under the CAC constraint. 
Rather, we follow the same steps of LP relaxation and obtain 
a scheme that stabilizes any $\vlambda$ for $\vlambda+\mbf{1} \in \mathring{\C}_{\Gamma}$, since
for large systems this is better than any arbitrarily close approximation
scheme (as $\vlambda \to \infty$ as $N \to \infty$).

\subsection{Inflexible System}
\label{sec:IDandIND}
\iffalse
We had obtained a capacity region (and optimal scheme) for IND systems and an outer region for ID systems
in Sec.~\ref{sec:capacity}. Now we present an alternate characterization of the capacity
region and outer region of inflexible systems, followed by an optimal centralized 
allocation scheme and a decentralized scheme with certain performance guarantees. 
This characterization is in terms of the bipartite graph $G=(V,E)$ that captures the
restriction of job-agent allocations.
\fi
Below we present a characterization of the capacity region of inflexible systems
in terms of the bipartite graph $G=(V,E)$, which captures the restriction of 
job-agent allocations.

\begin{theorem}
\label{thm:altStability}
Any $\vlambda$ can be stabilized if $\vlambda \in \mathring{\C}^I$, where
\[
\C^I = \left\{\vlambda =\sum_{l \in [L]} \vlambda^{(l)}: \ \vlambda^{(l)} \in \C^{(l)}_{\Gamma_l}, \lambda_j^l=0 
\ \mbox{ for all } (j,l) \notin E\right\}\mbox{,}
\]
where $\C^{(l)}_{\Gamma_l}$ is the capacity region of a single category system with an agent
availability distribution $\Gamma_l=\Gamma\left(U^l_i: i \in [M^l]\right)$. 
Moreover, no $\vlambda \notin \bar{\C}^I$ can be stabilized, i.e., for inflexible
systems the capacity region $\C_{\Gamma}=\bar{\C}^I$.
\end{theorem}
\begin{IEEEproof}
See Appendix II.\ref{sec:thm:altStability}.
\end{IEEEproof}

This theorem has the following simple consequence.
Consider separate {\em pools} of agents for each different category, cf.~\cite{VarshneyACSOLR2014_arXiv}, which has
agent-availability distributions $\{\Gamma_l: l \in [L]\}$. Each such pool (category) of agents $l$
can stabilize job-arrival rates in $\mathring{\C}^{(l)}_{\Gamma_l}$. Thus if the job arrival process of each
job type $j$ can be split in such a way that pool (category) $l$ of agents sees an arrival rate
$\lambda_j^l$, where $\lambda_j^l>0$ only if $(j,l) \in E$, while ensuring that
$\{\lambda_j^l:j\} \in \mathring{\C}^{(l)}_{\Gamma_l}$, the system would be stable. 

In a server farm  where jobs can be placed on any of the 
server queues, the join-shortest-queue (JSQ) policy stabilizes any stabilizable 
rate \cite{srikantYing2014}. JSQ gives an arriving job to the server with the shortest 
queue and each server serves jobs in FIFO order. For multi-category crowd systems, we can draw a
parallel between servers and agent pools. In addition we have constraints on job
placement given by $G$ and also have to do allocations of jobs among the agents
in the pool optimally (unlike JSQ we do not have FIFO/LIFO specified). 
Thus we have to adapt JSQ appropriately based on our insights about optimal 
operation of crowd systems.

We propose a statistics-agnostic scheme, JLTT-MWTA (Alg.~4) that has two parts:
JLTT (join least total task) directs arrivals to appropriate pools of agents 
and MWTA allocates jobs in each pool separately. 
Letting $Q^l_{j,s}(t)$ be the number of unallocated $(j,s)$-tasks in $l$th pool
just after epoch $t-1$, JLTT uses these quantities to direct jobs to 
appropriate pools whereas MWTA uses them to allocate tasks within each pool.

\begin{algorithm}
\label{alg:JSQandSRMW}
\caption{JLTT-MWTA:  Divide and Allocate}
\begin{algorithmic}
\STATE Input: $\mbf{A}(t)$, $\mbf{U}(t)$, $\mbf{Q}(t)$
\STATE Output: Job division and allocation
\STATE Create pool $l$ with category $l$ agents ($\forall l \in [L]$)
\STATE JLTT: Join Least Total Task 
\FOR{each $(j,s)$}
\STATE \hspace{0.1 in} Count number of unallocated $(j,s)$-tasks in pool $l$: $Q^l_{j,s}(t)$
\STATE \hspace{0.1 in} Divide $A_j(t) 1({r_{j,s}>0})$ 
tasks equally among pools $\arg \min_{l: (j,l) \in E} \sum_s Q^l_{j,s}(t)$
\ENDFOR
\STATE In each pool $l$ run MWTA for single-category system
\end{algorithmic}
\end{algorithm}

The JLTT part is computationally light. The central controller only needs to know $\mbf{Q}(t)$
and has to pick the minimally loaded ($\min_l \sum_s Q^l_{j,s}$)
pools of agents to direct jobs (type $j$). To perform MWTA in each pool, 
a PTAS, FPTAS, or LP relaxation scheme can be used.

Unlike JSQ, where service discipline in each server is fixed and the goal is to place
the jobs optimally, we have jobs with multi-dimensional service requirements from
time-varying stochastic servers (agent-availability) 
and have to place jobs as well as discipline the service in each random
and time-varying virtual pool. Thus optimality of JSQ cannot be claimed in our case.
But as stated below, JLTT division followed by MWTA allocation is indeed optimal.

\begin{theorem}
\label{thm:JSQandSRMW}
JLTT-MWTA stabilizes any $\vlambda \in \mathring{\C}^I$.
\end{theorem}
\begin{IEEEproof}
See Appendix~II.\ref{sec:thm:JSQandSRMW}.
\end{IEEEproof}
An important aspect of JLTT-MWTA is 
that job allocations within each pool can happen independently of
each other. The central controller only has to make a decision on how to split the jobs based on 
the current system state information. This allows a more distributed allocation along
the lines of following hierarchical organizational structure \cite{AgarwalKPS2011}.
First, the central controller divides jobs for different agent-pools based on $\{Q^l_{j,s}\}$.
Then in each agent pool, allocations are according to GreedyJob allocation, which works for
both decomposable and non-decomposable single-category systems. The distributed scheme that 
we propose here is an improvisation of the above JLTT scheme followed by GreedyJob allocation
in each pool. We call it Improvised JLTT and GreedyJob Allocation (Alg.~5).

\begin{algorithm}
\label{alg:ImprJSQandSRMW}
\caption{Improvised JLTT and GreedyJob Allocation}
\begin{algorithmic}
\STATE Input: $\mbf{A}(t)$, $\mbf{U}(t)$, $\mbf{Q}(t)$
\STATE Output: Job division and allocation
\STATE {\bf Improvised JSQ for each job-type $j$ and each skill $s$:}
\STATE \hspace{0.1 in} $n=0$
\STATE \hspace{0.1 in} $N^l_{j}=Q^l_{j}(t)$
\STATE \hspace{0.1 in} {\bf While}  \ \   $n<A_j(t)$
\STATE \hspace{0.2 in} Send $1$ task to the category $l^*$ with lowest \\
\STATE \hspace{0.2 in} index among $\arg \min_{l: (j,l) \in E} N^l_{j}$
\STATE \hspace{0.2 in} Increase $N^{l^*}_{j}$ and $n$ each by $1$
\STATE \hspace{0.1 in} {\bf End While} 
\STATE {\bf Allocations within each pool $l$:} 
\STATE \hspace{0.1 in}   Run GreedyJob allocation

\end{algorithmic}
\end{algorithm}

First note that unlike JLTT-MWTA, here we only maintain number of unallocated jobs and
do not maintain number of unallocated tasks for each skill $s$. This is because as 
GreedyJob allocation is used as allocation scheme in each pool, 
$Q^l_{j,s}=Q^l_{j,s'}$ for all $s,s'$. 

This algorithm is also simple to implement. The central controller only needs to
track the number of unallocated jobs ($Q^l_{j}(t)$) 
from the previous epoch and set $N^l_{j}=Q^l_{j}$.
For any arriving job of type $j$, the central controller sends the
job to the pool with minimum $N^l_{j}$ and updates $N^l_{j}$. This continues until the
next epoch, when the $N^l_{j}$ are reset to new $Q^l_{j}$ values.

Recall that Sec.~\ref{sec:intro} gave a simple example of a
fully distributed scheme where jobs pick agents greedily (from the set of feasible agents as per
$G$) and showed it was not a good scheme. Improvisation of JLTT is 
proposed for a better performance guarantee, while GreedyJob in each pool is proposed
for implementability and freedom of selection for customers. It is not hard to prove Improvised JLTT
followed by MWTA is optimal for any arrival and availability process 
satisfying the assumptions of Sec.~\ref{sec:model}. 
Below we present performance guarantee for Improvised JLTT and GreedyJob 
allocation.

To present performance guarantees of the distributed scheme 
we give an outer region $\C^{O}$ for the system,
along the lines of the alternative characterization $\C^I$ of
the capacity region $\C_{\Gamma}$ for inflexible systems.
\begin{theorem}
\label{thm:altOuter}
Inflexible crowd systems cannot be stabilized for $\vlambda \notin \C^{{O}}$, where
$
\C^{{O}} = \left\{\vlambda: \vlambda=\sum_{l \in [L]} \lambda^{(l)} \ \mbox{where}\ \lambda^{(l)} \in \C^{out}_{\bm{\mu}^l}\right\}\mbox{,}
$
and $\C^{out}_{\bm{\mu}^l}$ is the outer region for the single category system comprising the $l$th 
category (pool) of agents with $\bm{\mu}^l=\EX\left[\mbf{U}^l\right]$.
\end{theorem}
\begin{IEEEproof}
See Appendix II.\ref{sec:thm:altOuter}.
\end{IEEEproof}
For job allocation in server farms, extant performance guarantees
are mostly for symmetric load, i.e., symmetric (almost) 
service and job arrival rates and regular graphs, cf.~\cite{tsitsiklis2013queueing, bramson2010randomized}. 
Unlike server farms, symmetric load (in terms of skill-hours) is not guaranteed in crowd systems 
by symmetric arrival rates and graphs.
This is because different types of jobs have different skill and hour requirements.
The following guarantee for crowd systems is for bounded asymmetry (sub-polynomial variation) in 
agent availability, complete graph, asymmetric job arrival rates, and asymmetric 
job requirements (extendable to regular graphs with additional assumptions
on symmetry of job arrival rates and requirements).
Note that because of the inflexibility constraint, a multi-category system with
a complete graph is not equivalent to a single-category system.

\begin{theorem}
\label{thm:ImprJSQSRMW}
Without loss of generality assume
the same ordering of agent types in each category, i.e., $M^l=M^{l'}=M/L$ and
$h^l_i=h^{l'}_i$ for all $l, l', i$. If the arrival processes $\{A_j(t)\}$ and the 
agent availability processes $\{{U}^l_i(t)\}$ are i.i.d.\ across time and independent 
across types (jobs and agents), all these processes are Gaussian-dominated 
(and/or Poisson-dominated), $\sum_i \max_{l,l'}|\mu^l_i-\mu^{l'}_i|$ and
$\max_{l,s,s'}|\sum_i \mu^l_i (h^l_{i,s} - h^l_{i,s'})|$ are $O\left(\mbox{subpoly}(N)\right)$
and $G$ is complete bipartite, then for any given $\alpha \in (0,1]$,
$\exists N_{\alpha}$ such that Improvised JLTT and Greedy-job stabilizes any 
$\vlambda \in (1-\alpha) \C^{O} := 
\left\{\vlambda: \frac{1}{1-\alpha} \vlambda \in \C^{O} \right\}$ and the maximum
number of unallocated jobs (across all types) is $O(\log N)$ with
probability $1-o\left(\frac{1}{N^2}\right)$.
\end{theorem}
\begin{IEEEproof}
See Appendix II.\ref{sec:thm:ImprJSQSRMW}.
\end{IEEEproof}
The proofs of Thm.~\ref{thm:greedyAgent}, \ref{thm:greedyJob}, and \ref{thm:ImprJSQSRMW} 
are all based on constructing queue-processes (different for the algorithms) 
that stochastically dominate the number of unallocated jobs, and 
bounding the steady state distributions of these processes 
using Loynes' construction and moment generating function techniques.

\subsection{Implementation of Decentralized Schemes on Crowd Platforms}

We have described the allocation schemes at the level of system abstraction
and discussed their performance. These schemes can be easily implemented on
crowd platforms as well. 

GreedyAgent allocation is completely decentralized, only requiring agents to abide by a norm for
picking partial tasks, which can be enforced by randomized vigilance and penalizing 
norm violators in reputation. If the payments
are the same, as it generally is in single-category systems where all jobs
require same quality, there is no incentive for agents to deviate from the norm.

Any arbitrary contention method among
agents will work for the algorithm, and hence the crowdsourcing platform only needs
to ensure that no two allocations are done simultaneously (as practiced in airlines
booking). Multiple allocations can also be allowed by the platform if they
do not conflict. Here the platform has to ensure that an agent
can place requests only for an amount of tasks it can actually perform,
given the constraints on available hours. Also, only one agent can request
for a task or a certain part of it. Once the agent has been declined,
it can place request(s) for task(s) of the same or lesser hours.
This can either be enforced by appropriate modification of the portals
by keeping tracks of total hours of requests placed or by vigilance.
GreedyJob can also be easily implemented on a crowd platform. The platform has to ensure
that jobs request agents and not the other way around. One way to implement this is
to allow jobs to place requests for agents while ensuring they do not request more than 
the required service. Also, the platform has to ensure that
skill-hour requests of no two jobs collide. This again can be ensured by serializing the
requests as above. Agents are expected to accept a requested task, as there is no difference between
tasks involving same skill since payments are the same. This can also be ensured by linking 
agent rating to rate of task-request acceptance.

It is apparent that GreedyJob offers choice to customers and 
GreedyAgent offers choice to agents. By allowing a customer (or an agent) to decline an approaching agent
(or a customer) request and to explore more options, only one option at a time, 
the platform can provide freedom of choice to agents and customers under both 
schemes while operating at capacity.

In case of multi-category systems, the platform only needs to direct
arriving jobs to the appropriate pool of agents, based on current backlog; the rest of the allocation 
happens as per GreedyJob. Directing a job to a category
of agents can be implemented in a crowd platform by making the job visible only to freelancers of that 
category and vice versa (similar to filtering done by search engines and online social networks)
or through explicit hierarchical organization into pools \cite{AgarwalKPS2011}.

\section{Beyond the Capacity Region}
\label{eq:jobDropping}
We have now characterized the capacity (and outer) regions of different classes of crowd systems, shown the existence of
computationally feasible centralized schemes that achieve these regions, and presented simple distributed schemes with minimal
centralized intervention and good performance guarantees for any arrival rate within the capacity region. 
In crowd systems, however, arrival rates may not be within the capacity region, since the platform may have little or no
control on resource (freelancer) planning, unlike traditional communication networks or cloud computing systems. Hence, an
important aspect of crowd systems is to turn down job requests. Indeed, deciding to decline a job must be done 
as soon as the job arrives because dropping a job after first being accepted adversely affects the reputation of the
crowd platform. 

Here, we propose a centralized scheme for a crowd system to decline jobs on arrival in a way 
that is fair across all job types. Our scheme is statistics-agnostic and works
even for independent but non-stationary arrival and availability processes.

We solve the following problem: given an arrival rate $\vlambda$, design a 
statistics-agnostic policy to accept $(1-\beta)\vlambda$ jobs on average and allocate them 
appropriately such that $\beta$ is the minimum for which the crowd system is stable. Note that 
if $\vlambda \in \mathring{\C}_{\Gamma}$, the minimum $\beta$ is $0$, else, it
is strictly positive. We want to design a statistics-agnostic policy without 
the knowledge of $\vlambda$ and $\C_{\Gamma}$.
As a benchmark, we consider the following problem for $\epsilon>0$,
when $\vlambda$ and $\C_{\Gamma}$ are known. 
\begin{eqnarray}
\min \beta \in [0,1] \quad \mbox{s.t.} \ (1-\beta) \vlambda + \epsilon \mbf{1} \in {\C}_{\Gamma}\mbox{.} \label{eq:static}
\end{eqnarray}
Given $\beta^*$, optimum of \eqref{eq:static},
$(1-\beta^*) \vlambda$ is within $\epsilon$ of the optimal rate of accepted jobs
for which the system is stabilizable.

As we want a scheme that is agnostic of $\vlambda$ and $\C_{\Gamma}$, 
we propose the following simple scheme, for $\nu>0$ and
$\tilde{Q}^l_{j,s}(t)$ is the number of unallocated accepted $(j,s)$-tasks
($(j,s)$-tasks directed to category $l$) in the system.
\begin{eqnarray}
& & \hspace{-0.1 in} \beta(t) = \arg \min_{\beta \in [0,1]} \nonumber \\
& & \hspace{-0.1 in} \begin{cases}
\beta \sum_j A_j(t) - \nu \beta  \sum_{j,s: r_{j,s}>0} \tilde{Q}_{j,s}(t) A_j(t) \hspace{0.4 in} \mbox{(I)} \\
\beta \sum_j A_j(t) - \nu \beta  \sum_{j,s: r_{j,s}>0} \min_l \tilde{Q}^l_{j,s}(t) A_j(t) \hspace{0.1 in}\mbox{(II)}
\end{cases} \nonumber \\
& & \hspace{-0.1 in}\mbox{Job is accepted w.p.} \ 1-\beta(t), \ \mbox{accepted jobs} \ \tilde{\mbf{A}}(t) \ (\mbox{I} \ \& \ \mbox{II})\nonumber \\
& & \hspace{-0.1 in}\mbox{For accepted jobs run} 
\begin{cases}
\mbox{MWTA} \hspace{0.3 in} \mbox{(I)} \\
\mbox{JLtT-MWTA} \hspace{0.1 in} \mbox{(II)}
\end{cases} \label{eq:Drop} 
\end{eqnarray}
Steps marked by I (II) are applicable for FD, FND and IND (ID and IND) systems.

\begin{theorem}
\label{thm:Drop}
Crowd system with jobs accepted and allocated according to \eqref{eq:Drop} is stable and
$\sum_j \lambda_j (1-\beta^*) - \frac{1}{T} \sum_{t=1}^T \EX[\sum_j \tilde{A}_j(t)]$ can be made 
arbitrarily small for an appropriately chosen $\nu$, for all sufficiently large $T$.
\end{theorem}
\begin{IEEEproof}
See Appendix II.\ref{sec:thm:Drop}.
\end{IEEEproof}

This theorem demonstrates that by following the job acceptance and allocation method
\eqref{eq:Drop}, the crowd system can be stabilized while ensuring the average number
of accepted jobs per allocation epoch is arbitrarily close to the optimal number of
accepted jobs per allocation period. Note that as all jobs (across all types)
are accepted with the same probability, the above result also implies that
$(1-\beta^*)\lambda_j-\frac{1}{T} \sum_{t=1}^T \EX[\tilde{A}_j(t)]$ is small.
It can be shown that the above scheme works
for time-varying systems ($\EX[\mbf{A}(t)]=\vlambda(t)$ and $\mbf{U}\sim \Gamma^t()$) 
as well guaranteeing small $\lim_{T \to \infty} \frac{1}{T} \sum_{t=1}^T\sum_j \left(\lambda_j(t) (1-\beta^*(t))-\EX[\tilde{A}_j(t)]\right)$, where $\beta^*(t)$ is the solution of (\ref{eq:static}) for 
$\vlambda=\vlambda(t)$ and $\C_{\Gamma}=\C_{\Gamma^t}$.

section{Conclusion}
\label{sec:conclusion}

Human information processing, structured through freelance markets, is an emerging
structure for performing informational work by capturing the cognitive energy of the crowd.  
It is important to understand the fundamental limits and optimal designs for such systems.

In this work we provide a characterization of the work capacity of 
crowd systems and present two statistic-agnostic job allocation schemes MWTA (flexible jobs) and JLTT-MWTA (inflexible jobs) to achieve limits. To ensure low computational
load on the crowd platform provider and freedom of choice for job requesters, we present
simple decentralized schemes, GreedyAgent, GreedyJob, and Improvised JLTT-GreedyJob
that (almost) achieve capacity with certain performance
guarantees. These decentralized schemes are easy to implement on crowd platforms, require
minimal centralized control, and offer freedom of self-selection to customers: all
desirable qualities for any crowd platform. Due to quick evolution and unpredictability of freelancer resources, crowd systems may often operate outside capacity, which inevitably results in huge 
backlogs. Backlogs hurt the reputation of the platform, and so we also propose a scheme that 
judiciously accepts or rejects jobs based on
the system load. This scheme is fair in accepting jobs across all types and accepts the
maximum number of jobs under which the system can be stable.

\bibliographystyle{IEEEtran}
\bibliography{abrv,conf_abrv,crowd}

\renewcommand{\appendixname}{Appendix I~}

\begin{appendix}
\label{sec:CompImpl}
\subsection{Computation for Centralized Allocation}
For a single category system ($L=1$), note that $z^1_{j,s}=a_{j,s} \in \Z_+$ and hence the
feasibility condition \eqref{eq:decompCond} becomes
\[
\sum_{j} a_{j,s} r_{j,s} \le \sum_i u_{i} h_{i,s} \mbox{ for all } s\in[S], a_{j,s}\in \Z_+  \mbox{,}
\]
with condition \eqref{eq:nonDecompCond} additionally requiring $a_{j,s}=a_{j,s'}$ for all
$j, s, s'$. Thus, $C(\bu)$ is the set of $\{a_{j,s}\}$ satisfying the above conditions
for respective classes of jobs (as well as systems) and $\C$ is the weighted (by $\Gamma(\bu)$)
sum of convex hulls of $C(\bu)$s.

$\C^{out}_{\bm{\mu}}$ has a simple characterization as well. 
As for any $j \in [N]$, $(j,1) \in E$, and $\mc{N}(J)=1$ for all $J \subset [N]$,
$\sum_{l \in \mc{N}(J)} \sum_{i \in [M^l]} \mu^{l}_i h^{l,i}_s = \sum_{i \in [M]} \mu_i h_{i,s}$.
Thus it is sufficient to satisfy the inequality for $J=[N]$, and hence,
\[
\C^{out}{\bm{\mu}} = \left\{ \vlambda: \sum_{j \in [N]} \lambda_j r_j \le  \sum_{i \in [M]} 
\mu_i h_{i}\right\} \mbox{.}
\]

The MaxWeight computation in MWTA for single-category 
decomposable systems turns out to be the 
following integer linear program (ILP), which is related to knapsack problems.
\begin{align}
& \arg \max_{\Delta_{j,s}:j,s} \sum_{j,s}   Q_{j,s} \Delta_{j,s} \notag \\ 
& \mbox{s.t.} \ \sum_{j} a_{j,s} r_{j,s} \le \sum_i u_{i} h_{i,s} \forall s\in[S], 
\label{eq:MWforSC} 
\end{align}

This problem is an integer program,
hence it is not clear whether this problem can be solved efficiently at all instants. 
In fact, it is a so-called unbounded knapsack problem for a given
$\bu$ and $\mbf{Q}$. This problem is known to be NP-hard \cite{kellerer2004knapsack}. There is a pseudo-polynomial algorithm based on dynamic
programming which solves it exactly, but the runtime may
depend on $Q_{j,s}$. This dynamic programming-based algorithm can be converted into a fully 
polynomial time approximation schemes (FPTAS) which achieves any $(1-\epsilon)$ approximation of
the problem in poly$\left(\frac{1}{\epsilon}\right)$ computations. Moreover, there 
exists faster greedy algorithms that achieve $\frac{1}{2}$ approximation and can be 
converted into a polynomial time approximation scheme (PTAS) that achieves $(1-\epsilon)$ 
approximation in poly$(n^\frac{1}{\epsilon})$ computations. Thus we can conclude that though 
the MWTA algorithm
is computationally hard for single-category decomposable system, there exist efficient approximation
schemes. It is not hard to show (Prop.~\ref{prop:approxCap} below) that an algorithm that gives $(1-\epsilon)$ 
approximation of the optimization problem in MWTA can stabilize any $\vlambda$ for which 
$\frac{\vlambda}{(1-\epsilon)} \in \C^D_{\Gamma}$.

For single-category non-decomposable systems, feasibility condition \eqref{eq:nonDecompCond}
of an allocation of $a_j$ jobs of type $j$ is given by,
\begin{align}
\sum_{j} a_{j} r_{j,s} & \le \sum_i u_{i} h_{i,s} \forall s\in[S], a_{j}\in \Z_+.  \nonumber
\end{align}
Hence the stabilizable region changes accordingly to
\begin{align}
& \C_{\Gamma} = \left\{\sum_{\bu} \Gamma(\bu) \lambda(\bu): \lambda(\bu) \in \C^{ND}(\bu)\right\}, \nonumber \\
& \C^{ND}(\bu) = \mbox{conv}\left\{a_{j,s}\in \Z_+: \sum_{j} a_{j} r_{j,s} \le \sum_i u_{i} h_{i,s} 
\ \mbox{and} \
a_{j,s}=a_j \forall s\right\}. \nonumber
\end{align}
where conv$\{\cdot\}$ is the convex hull. 

Hence, in this case, the MWTA allocation needs to solve
\begin{align}
& \arg \max_{\Delta_{j,s}:j} \sum_j \left(\sum_{s}   Q_{j,s} \Delta_{j,s}\right) \nonumber \\
& \mbox{s.t.} \ \sum_{j} \Delta_{j,s} r_{j,s} \le \sum_i u_{i} h_{i,s}, \Delta_{j,s} = \Delta_{j,s'},
\mbox{ for all } s, s' \in[S]\mbox{,}
\label{eq:MWnonDecompSC} 
\end{align}
and then divide the allocated jobs arbitrarily among agents while meeting their
per skill time-availability, as there is only one category of agents.

This problem is also a knapsack-like integer program, with an additional constraint that the number 
of $(j,s)$-items has to be the same as the number of 
$(j,s')$-items for all $j, s$ and $s'$. Such a problem is called a multi-dimensional knapsack problem. This problem is  also NP-hard. Moreover, provably there cannot exist a fully polynomial time
approximation scheme for this problem \cite{kellerer2004knapsack}.

For a multi-dimensional knapsack problem, there
exists an approximation scheme that achieves an approximation factor equal to the dimension $d$ 
\cite{kellerer2004knapsack}.
This approximation scheme can be converted into a PTAS with complexity $O\left(N^{\kappa}\right)$ 
and an approximation factor of $1-\epsilon$, where $\kappa$ is strictly increasing with dimension
and $\frac{1}{\epsilon}$. In the case of our setting, the number of skills $S$ (dimension) is large and may 
scale with $N$, hence this scheme is not suitable.

Though the total number of skills $S$ can scale with $N$, in most cases the number of skill-parts that a 
type $j$ job has is a constant, i.e., $r_{j}$ has most coordinates as $0$. Thus it is
apparent from the objective function that the optimal choice of $\Delta_{j,s}$ is $0$ for 
the corresponding coordinates $s$. This allows us to rewrite the optimization problem as another
multi-dimensional knapsack problem with constant dimensions given by $\max_j|\{s: r_{j,s}>0\}|$.

For this problem we can use the PTAS to obtain arbitrarily close approximation and hence can
stabilize a rate-region arbitrarily close to $\C^{ND}$. But the complexity of this algorithm is very
high as complexity scales super-exponentially ($N^k$) with the approximation factor (unlike  
decomposable systems where we have an FPTAS, polynomial in $\frac{1}{\epsilon}$).

Note that our goal is not to solve \eqref{eq:MWnonDecompSC} 
optimally, but to have a fast allocation scheme that can
stabilize a large fraction of $\C^{ND}$. In this regard we can take  
a different approach that exploits basic characteristics of a crowd system. 
Since $N$ and $\lambda_j(N)$ for $j \in [N]$ are large in most crowd systems, 
if an allocation scheme stabilizes
any rate $\vlambda$ for $\vlambda + c \mathbf{1} \in \C$, then it
stabilizes $\left(1-\max_i \frac{1}{\lambda_i(N)}\right)\C$. Note that
as $\lambda_i(N)$ scales with $N$, this implies that such a scheme
would stabilize almost all of $\C$.
Motivated by this, we propose the following allocation scheme, which is a
modification of \eqref{eq:MWnonDecompSC}.

\begin{align}
& \{\hat{x}_{j}, j\} = \arg \max_{x_{j} \in \R:j} \sum_j \left(\sum_{s}   Q_{j,s}\right) 
x_{j}\nonumber \\
& \mbox{s.t.} \ \sum_{j} x_{j} r_{j,s} \le \sum_i u_{i} h_{i,s},
\mbox{ for all } s\mbox{,}
\label{eq:MWnonDecompSCLP} 
\end{align}
and allocate $\tilde{\Delta}_{j}=\lfloor \hat{x}_{j}\rfloor$ jobs of type $j$ to the agents, 
splitting arbitrarily while meeting time-availability constraints. 

Note that since in \eqref{eq:MWnonDecompSCLP}, the variables are relaxed to $\R$ from $\Z$,
$\sum_j \left(\sum_{s}   Q_{j,s}\right) \hat{x}_{j} \ge \sum_j 
\left(\sum_{s}   Q_{j,s}\right) \hat{z}_{j}$. 
Again, $\tilde{\Delta}_{j}=\lfloor \hat{x}_{j}\rfloor \ge \hat{x}_{j} - 1$, hence
$\sum_j \left(\sum_{s}   Q_{j,s}\right) \tilde{\Delta}_{j} \ge \sum_j 
\left(\sum_{s}   Q_{j,s}\right) \left(\hat{z}_{j}-1\right)$. 

The following proposition guarantees that a proposed LP-relaxation 
scheme stabilizes any $\vlambda$ with $\lambda+\mathbf{1} \in \C^{ND}$.
\begin{proposition}
\label{prop:approxCap}
Let $\mathcal{P}$ be an allocation scheme that at epoch $t$ does an allocation 
$\{\Delta(t)\}$ instead of $\{\hat{\Delta}(t)\}$ of the MWTA allocation scheme, which satisfies 
\begin{align}
\sum_{j,s}  Q_{j,s}(t) \hat{\Delta}_{j,s}(t) \le \sum_{j,s}  Q_{j,s}(t) {\Delta}_{j,s}(t)
+ \sum_{j,s}  Q_{j,s}(t) \delta, \nonumber
\end{align}
or,
\begin{align}
(1-\epsilon)\sum_{j,s}  Q_{j,s}(t) \hat{\Delta}_{j,s}(t) \le \sum_{j,s}  Q_{j,s}(t) {\Delta}_{j,s}(t), \nonumber
\end{align}
stabilizes any rate $\vlambda \in \C$ if $\vlambda+\delta\mbf{1} \in \C$ or $\frac{1}{1-\epsilon}\vlambda \in \C$ respectively.
\end{proposition}
\begin{IEEEproof}
This proof follows the same steps as the proof of Thm.~\ref{thm:SRMW}.
We first prove the result for an allocation with
\begin{align}
\sum_{j,s} Q_{j,s}(t) \hat{\Delta}_{j,s}(t) \le \sum_{j,s}  Q_{j,s}(t) {\Delta}_{j,s}(t)
+ \sum_{j,s}  Q_{j,s}(t) \delta\mbox{.} \nonumber
\end{align}
as follows:
\begin{align}
&\EX\left[L\left(\mbf{Q}(t+1)\right) - L\left(\mbf{Q}(t)\right)|\mbf{Q}(t)\right] \nonumber \\
&\quad\le \EX\left[\sum_{j,s} Q_{j,s}(t) A_j(t)|\mbf{Q}(t)\right] - \EX\left[\sum_{j,s}  Q_{j,s}(t) \left(\hat{\Delta}_{j,s}(t)-\delta\right)|\mbf{Q}(t)\right] + \EX\left[\sum_{j,s} \left(A^2_j(t)+(\hat{\Delta}_{j,s}-\delta)^2\right)|\mbf{Q}(t)\right] 
\nonumber
\end{align}
We can bound the last term as above, 
because $(\hat{\Delta}_{j,s}-\delta)^2 \le \hat{\Delta}^2_{j,s}+\delta^2$.
Thus,
\[
\EX\left[L\left(\mbf{Q}(t+1)\right) - L\left(\mbf{Q}(t)\right)|\mbf{Q}(t)\right] \le B + \EX\left[\sum_{j,s} Q_{j,s}(t) (A_j(t)+\delta)|\mbf{Q}(t)\right] - \EX\left[\sum_{j,s} Q_{j,s}(t) \hat{\Delta}_{j,s}(t)|\mbf{Q}(t)\right]
\]

Now note that if $\vlambda$ is such that $\vlambda+\delta\mbf{1} \in \C$, then we can write it in terms of convex combinations of $\mbf{d} \in C(\bu)$ and follow the same steps as 
in the proof of Thm.~\ref{thm:SRMW}.

Similarly for the other case of constant factor approximation we have:
\[
\EX\left[L\left(\mbf{Q}(t+1)\right) - L\left(\mbf{Q}(t)\right)|\mbf{Q}(t)\right] \le B + \EX\left[\sum_{j,s}  Q_{j,s}(t) A_j(t)|\mbf{Q}(t)\right] - \EX\left[\sum_{j,s}  Q_{j,s}(t) \hat{\Delta}_{j,s}(t)(1-\epsilon)|\mbf{Q}(t)\right] 
\]
as $(1-\epsilon)^2 \hat{\Delta}^2_{j,s} \le \hat{\Delta}^2_{j,s}$.

If $\vlambda \in (1-\epsilon) \C$, then by definition of $\C$ and $(1-\epsilon)\C$,
there exist $\nu(\bu) \in \C(\bu)$ and $\{\mbf{d}_k(\bu)\in C(\bu)\} $ such that, 
$\vlambda \le (1-\epsilon) \nu(\bu) \Gamma(u)$  and $\sum_k \gamma_k \mbf{d}^k = \nu(\bu)$
for $\gamma_k>0, \sum_k \gamma_k = 1$. As for any $\mbf{d}^k$ for a given $\bu$,
$\sum_{j,s} Q_{j,s}(t) \hat{\Delta}_{j,s}(t) \le \sum_{j,s} Q_{j,s}(t) \mbf{d}^k_{j,s}$,
$\sum_{j,s} Q_{j,s}(t) \lambda_j \le \EX\left[\sum_{j,s}  Q_{j,s}(t) \hat{\Delta}_{j,s}(t)|\mbf{Q}(t)\right]$. Then following the proof of Thm.~\ref{thm:SRMW},
the result follows.
\end{IEEEproof}
\end{appendix}

\renewcommand{\appendixname}{Appendix II~}
\begin{appendix}

\label{sec:proofs}
In this appendix, we present proofs of the main results in 
Secs.~\ref{sec:capacity}--\ref{sec:multiCategory}.
As mentioned earlier, most of these results extend to systems with
stationary and ergodic arrival and availability processes, but here 
we only present results for i.i.d.\ processes.

\subsection{Proof of Theorem \ref{thm:capacityGen}}
\label{sec:thm:capacityGen}

Here we only prove the converse part, i.e., $\vlambda \notin \C$ cannot be
stabilized by any policy. For the direct part, it is sufficient to prove
there exists a scheme that stabilizes any $\vlambda \in \C$, and so
the proof of Thm.~\ref{thm:SRMW} below is sufficient.

First we prove that $\C$ is a convex subset of $\R_+^N$.
If $\vlambda, \vlambda' \in \C$, then there exist
$\left(\vlambda(\bu) \in \C(\bu): \bu \in \Z_+^{M}\right)$ and 
$\left(\vlambda'(\bu)\in \C(\bu): \bu \in \Z_+^{M}\right)$ such that
\[
\sum_{\bu} \Gamma(\bu) \vlambda(\bu) = \vlambda\mbox{,}\quad
\sum_{\bu} \Gamma(\bu) \vlambda'(\bu) = \vlambda'\mbox{.}
\]

Thus for any $\gamma \in [0,1]$, 
\begin{align*}
\gamma \vlambda + (1-\gamma) \vlambda' = \sum_{\bu} \Gamma(\bu) (\gamma \vlambda(\bu) + (1-\gamma) \vlambda'(\bu).
\end{align*}
Note that $\C(\bu)$ is convex since it is the convex hull of $C(\bu)$; hence $\gamma \vlambda(\bu) + (1-\gamma) \vlambda'(\bu) \in \C(\bu)$, which in turn implies $\gamma \vlambda + (1-\gamma) \vlambda' \in \C$. This proves convexity of $\C$.

Thus $\bar{\C}$ is a closed convex set. Hence for any $\vlambda^O \notin \bar{\C}$, there exists a
hyperplane $h^Tx=c$ that separates $\bar{\C}$ and $\vlambda^O$, i.e., there exists an $\epsilon>0$
for any $\vlambda \in \bar{\C}$ such that $h^T \vlambda^O \ge h^T \vlambda + \epsilon$. 

Hence under any policy: 
\begin{align}
\EX\left[h^T \mbf{Q}(t+1)\right] 
& = \EX\left[h^T \left(\mbf{Q}(t)+\mbf{A}(t)-\mbf{D}(t)\right)\right] \nonumber \\
& = \EX\left[h^T \left|\mbf{Q}(t)+\mbf{A}(t)-\mbf{\Delta}(t)\right|^{+}\right] \nonumber 
\end{align}
where $\mbf{\Delta}(t)$ is the number of possible departure under the scheme if there were infinite
number of jobs of each type, and $|\cdot|^{+}$ is shorthand for $\max(\cdot,0)$. As $|x|^{+}$ is a convex function of $x$, 
$h^T \left|\mbf{Q}(t)+\mbf{A}(t)-\mbf{\Delta}(t)\right|^{+}$ is a convex function of
$\mbf{Q}(t), \mbf{A}(t)$, and $\mbf{\Delta}(t)$. Thus by Jensen's inequality:
\[
\EX\left[h^T \left|\mbf{Q}(t)+\mbf{A}(t)-\mbf{\Delta}(t)\right|^{+}\right] \ge h^T \left|\EX\left[\mbf{Q}(t)\right]+\EX\left[\mbf{A}(t)\right]-\EX\left[\mbf{\Delta}(t)\right]\right|^{+}
\]

Note that any $\vlambda$ is a $\Gamma(\bu)$-combination of some 
$\{\lambda(\bu) \in \C(\bu)\}$ and any $\lambda(\bu)$ is some convex combination of elements
of $C(\bu)$. Also, from the allocation constraints it is apparent that if $\mbf{a} \in C(\bu)$ then
also $\mbf{a}' \in C(\bu)$ if $\mbf{a}' \le \mbf{a}$. These two imply that 
for any $\vlambda \in \bar{\C}$, if there exists
a $\vlambda' \le \vlambda$ (component-wise) and $\vlambda'\ge \mbf{0}$, then $\vlambda' \in \bar{\C}$.
That is $\C$ is {\em coordinate convex}.
This in turn implies that for any $\lambda^O \notin \bar{\C}$  there exists an $h\neq\mbf{0} \in \R_+^N$ 
such that hyperplane separation holds for this $h$. Thus for $h \ge \mbf{0}$:
\begin{align}
\EX\left[h^T \mbf{Q}(t+1)\right] &\quad\ge h^T \left|\EX\left[\mbf{Q}(t)\right]+\EX\left[\mbf{A}(t)\right]-\EX\left[\mbf{\Delta}(t)\right]\right|^{+} \nonumber \\
&\quad= \sum_j \left|h_j \EX\left[{Q}_j(t)\right] + h_j\EX\left[{A}_j(t)\right] - h_j\EX\left[{\Delta}_j(t)\right]\right|^{+} \nonumber  \\
&\quad\ge \sum_j \left(h_j \EX\left[{Q}_j(t)\right] + h_j\EX\left[{A}_j(t)\right] - h_j\EX\left[{\Delta}_j(t)\right]\right) \nonumber \\
&\quad\ge \EX\left[h^T \mbf{Q}(t)\right] + h^T \EX\left[{\vlambda^O}\right] - \sup_{\vlambda \in \bar{\C}} 
h^T \vlambda \nonumber \\ 
&\quad\ge \EX\left[h^T \mbf{Q}(t)\right] + \epsilon \nonumber
\end{align}
Thus we have $\EX\left[h^T \mbf{Q}(t+1)\right] \to \infty$. As $h\ge 0$, this implies 
there exists $j$ such that $\EX[Q_j(t)] \to \infty$ as $t \to \infty$. 
Hence, the system is not stable.

\subsection{Proof of Theorem \ref{thm:outerGen}}
\label{sec:thm:outerGen}
Consider the dynamics of $Q_{j,s}(t)$, the unallocated $(j,s)$ tasks at the end of epoch
$t$. 
\begin{align}
Q_{j,s}(t+1) &= \left|Q_{j,s}(t) + A_j(t) - D_{j,s}(t)\right|^{+} \nonumber \\
              &\ge Q_{j,s}(t) + A_j(t) - D_{j,s}(t) \nonumber \\
              &\ge \sum_{k=0}^t \left(A_j(t) - D_{j,s}(t)\right)\mbox{.} \nonumber
\end{align}

As $r_{j,s} \ge 0$,
\begin{align}
\EX\left[r_{j,s} Q_{j,s}(t+1)\right] & \ge \sum_{k=0}^t \EX\left[A_j(t) r_{j,s} - r_{j,s} D_{j,s}(t)\right] \nonumber \\
                                     & = \sum_{k=0}^t \left(\lambda_j r_{j,s} - r_{j,s} \EX\left[D_{j,s}(t)\right] \right) \nonumber
\end{align}

Consider any set $J \subset [N]$, then at any epoch $t$, to schedule a certain number of tasks
of type $(j,s)$, the system needs that much available usable skill-hours. This follows from
conditions \eqref{eq:decompCond} and \eqref{eq:nonDecompCond} and can be written as:
\begin{align*}
\sum_{j \in J} r_{j,s} D_{j,s}(t) \le \sum_{l \in \mathcal{N}(J)} \sum_{i \in [M^l]} h^l_{i,s} U^l_{i,s}\mbox{.}
\end{align*}
This in turn implies
\[
\sum_{j \in J} r_{j,s} \EX\left[D_{j,s}(t)\right]  \le \sum_{l \in \mathcal{N}(J)} \sum_{i \in [M^l]} h^l_{i,s} \mu^l_{i,s}\mbox{.}
\]
Hence,
\begin{align}
\EX\left[\sum_{j \in J} r_{j,s} Q_{j,s}(t+1)\right] &\ge \sum_{k=0}^t \left(\sum_{j \in J}  \lambda_j r_{j,s} - \sum_{j \in J} r_{j,s} \EX\left[D_{j,s}(t)\right] \right)  \nonumber \\
&\ge \sum_{k=0}^t \left(\sum_{j \in J}  \lambda_j r_{j,s} - \sum_{l \in \mathcal{N}(J)} \sum_{i \in [M^l]} h^l_{i,s} \mu^l_{i,s}\right)
\end{align}
For any $\vlambda \notin \bar{\C}^{out}$, by definition there exists a $J \subset [N]$ such that
$\sum_{j \in J}  \lambda_j r_{j,s} - \sum_{l \in \mathcal{N}(J)} \sum_{i \in [M^l]} h^l_{i,s} \mu^l_{i,s} > 0$. Thus in that case, $\lim \sup_{t\to \infty} \EX\left[\sum_{j \in J} r_{j,s} Q_{j,s}(t+1)\right] = 
\infty$. Note that since $J$ is finite and so is $\max_j r_{j,s}$, there exists a $j \in J$ such that
$\lim \sup_{t\to \infty} \EX\left[Q_{j,s}(t+1)\right] = \infty$. This shows the system is not stable
for $\vlambda \notin \bar{\C}^{out}$ and proves $\C_{\Gamma} \subset \C^{out}$.

\subsection{Proof of Proposition \ref{prop:TA}}
\label{sec:prop:TA}
We consider FD, FND, and IND cases separately. 
\begin{itemize}
\item[{\em FD:}] The MaxWeight part chooses $z^l_{j,s}$ to be integral which
implies that integral number of tasks can be allocated if done appropriately.
As hours are allocated from tasks in order, a task later in the order only gets
allocated (partially or fully) after the tasks before it are fully allocated.
This leads to no partially-allocated tasks.

\item[{\em FND:}] Same ordering is used for all $(j,s)$ tasks and MaxWeight
chooses $z^l_{j,s}$ such that $a_{j,s} = a_{j,s'}$ (as it satisfies FND) and 
hence if an $s$-task of a job is chosen then also $s'$ is chosen for 
$r_{j,s}, r_{j,s'}>0$.

\item[{\em IND:}] Same ordering is used for all $(j,s)$ tasks, allocations
to different categories are in same order ($l=1$ to $L$) and MaxWeight
chooses $z^l_{j,s}$ such that $z^l_{j,s} = z^l_{j,s'}$ for all $l,l'$ (as it
satisfies IND), hence if a task of a job is allocated to category 
$l$ then so are the other tasks.
\end{itemize}

\subsection{Proof of Theorem \ref{thm:SRMW}}
\label{sec:thm:SRMW}
Note that MaxWeight chooses an allocation $\{\hat{\Delta}_{j,s}(t)\}$. But, the maximum number of $(j,s)$-tasks
that can be served is $Q_{j,s}(t)+A_j(t)$. By Prop.~\ref{prop:TA}, Task Allocation does a feasible allocation
for FD, FND, and IND systems. Also, note that in the Task Allocation algorithm, the number of allocated $(j,s)$-tasks is
$\hat{D}_{j,s}=\min \left(\hat{\Delta}_{j,s}(t),Q_{j,s}(t)+A_j(t)\right)$.

Consider the usual Lyapunov function $L\left(\mbf{Q}\right)=\sum_{j,s} Q^2_{j,s}$.
\begin{align*}
&\EX\left[L\left(\mbf{Q}(t+1)\right) - L\left(\mbf{Q}(t)\right)|\mbf{Q}(t)\right] \nonumber \\
&\quad =\EX\left[\sum_{j,s}  \left(Q^2_{j,s}(t+1) - Q^2_{j,s}(t)\right)|\mbf{Q}(t)\right] \nonumber \\
&\quad =\EX\left[\sum_{j,s}  \left(\left(Q_{j,s}(t)+A_j(t)-\hat{D}_{j,s}\right)^2 - Q^2_{j,s}(t)\right)|\mbf{Q}(t)\right]  \\
&\quad \le \EX\left[\sum_{j,s}  \left(\left(Q_{j,s}(t)+A_j(t)-\hat{\Delta}_{j,s}\right)^2 - Q^2_{j,s}(t)\right)|\mbf{Q}(t)\right] \\
&\quad \le \EX\left[\sum_{j,s}  Q_{j,s}(t) A_j(t)|\mbf{Q}(t)\right] - \EX\left[\sum_{j,s}  Q_{j,s}(t) \hat{\Delta}_{j,s}(t)|\mbf{Q}(t)\right] + \EX\left[\sum_{j,s} \left(A^2_j(t)+\hat{\Delta}^2_{j,s}\right)|\mbf{Q}(t)\right] 
\end{align*}
We bound the last term first.
\begin{align}
\EX\left[\sum_{j,s} \left(A^2_j(t)+\hat{\Delta}^2_{j,s}\right)|\mbf{Q}(t)\right] &= \sum_{j,s}  \EX[A^2_j] + \EX\left[\sum_{j,s} \left(r_{j,s} \hat{\Delta}_{j,s}\right)^2\right]
\nonumber \\
& \le \sum_{j,s} \EX[A^2_j]  + \frac{1}{\min(r_{j,s}>0)}\EX\left[\left(\sum_{j,s} r_{j,s} \hat{\Delta}_{j,s}\right)^2\right] \nonumber \\
& \le \sum_{j,s} \EX[A^2_j] + \frac{1}{\min(r_{j,s}>0)} \EX\left[\left(\sum_{i,l} \left(\sum_s h^l_{i,s}\right) U^l_{i}\right)^2\right] \label{eq:SRMW4} \\
& \le \sum_{j,s} \EX[A^2_j] + \frac{\left(\max_{l,i,s} h^l_{i,s}\right)^2}{\min(r_{j,s}>0)} \max_{l,i} 
\EX\left[\left(U^l_i\right)^2\right] \frac{M(M+1)}{2} \label{eq:SRMW5}
\end{align}
This is a constant $B<\infty$ independent of $\mbf{Q}$, 
as $\EX[A^2_j]$ and  $\EX\left[\left(U^l_i\right)^2\right]$ are finite for all $j,l,i$.

To bound the first term, note that if $\lambda+\epsilon \mbf{1} \in \C$, then there 
exist $\{\nu(\bu) \in \C(\bu)\}$ such that $\lambda_j \le \sum_{\bu} \Gamma(\bu) \nu(\bu) - \epsilon$
for all $j \in [J]$. Again note that as $\C(\bu)$ is the convex hull of 
$C(\bu)$, $\nu(\bu)=\sum_{k} \gamma_k \mbf{d}^k(\bu)$ for some $\{\mbf{d}^k(\bu) \in C(\bu)\}$ and $\gamma_k \ge 0$
with $\sum_k \gamma_k = 1$. So,
\begin{align*}
\EX\left[\sum_{j,s} Q_{j,s}(t)  A_j(t)|\mbf{Q}(t)\right] &= \sum_{j,s} Q_{j,s}(t)  \lambda_j \\
& \le \sum_{j,s} Q_{j,s}(t)\nu_j-\epsilon\sum_{j,s} Q_{j,s} \\
& \le \sum_{j,s} Q_{j,s}(t) \sum_{\bu} \Gamma(\bu) \sum_{k} \gamma_k d^k_j(\bu) -\epsilon\sum_{j,s}  Q_{j,s} \\
&=\sum_{\bu} \Gamma(\bu) \sum_{j,s} Q_{j,s}(t) \sum_{k} \gamma_k d^k_j(\bu) - \epsilon\sum_{j,s}  Q_{j,s} \nonumber \\
& \le \sum_{\bu} \Gamma(\bu) \max_{\mbf{d}(\bu)\in C(\bu)}\sum_{j,s} Q_{j,s}(t) d_{j,s}(\bu) - \epsilon\sum_{j,s}  Q_{j,s} \nonumber \\
& = \EX\left[\sum_{j,s}  Q_{j,s}(t) \hat{\Delta}_{j,s}(t)|\mbf{Q}(t)\right] - 
\epsilon\sum_{j,s} Q_{j,s}. 
\end{align*} 

Thus, we have a bound on the Lyapunov drift,
\[
\EX\left[L\left(\mbf{Q}(t+1)\right) - L\left(\mbf{Q}(t)\right)|\mbf{Q}(t)\right]  \le B - \epsilon\sum_{j,s} Q_{j,s} \mbox{.}
\]
Hence,
\[
\EX\left[L\left(\mbf{Q}(T)\right) - L\left(\mbf{Q}(0)\right)\right] \le B T - \epsilon\sum_{t=0}^{T-1} \sum_{j,s} w_{j,s} \EX[Q_{j,s}(t)] \mbox{.}
\]
As $L(\mbf{Q}(0))<\infty$ and $L(\mbf{Q})\ge 0$ for all $\mbf{Q}$, we have that for all $T$,
\[
\frac{1}{T} \sum_{t=0}^{T-1} \sum_{j,s} \EX[Q_{j,s}(t)] \le \frac{B}{\epsilon} + \frac{L(0)}{T}
< \infty 
\]
This in turn implies $\lim \sup_{t \to \infty} \sum_{j,s} \EX[Q_{j,s}(t)] < \infty$,
otherwise the time-average cannot be finite. This implies that for all $(j,s)$ with $r_{j,s}>0$,
$\lim \sup_{t \to \infty}  \EX[Q_{j,s}(t)] < \infty$. 

Again note that $Q_j(t) \le \sum_{s} Q_{j,s}$, as there can be unallocated jobs with more than
one part unallocated. Hence, $\lim \sup_{t \to \infty}  \EX[Q_{j}(t)] < \infty$ for all $j \in [N]$.

\subsection{Proof of Theorem \ref{thm:greedyAgent}}
\label{sec:thm:greedyAgent}
In the GreedyAgent algorithm, as each agent with available skill-hours greedily chooses
to serve a task, no $(j,s)$ task of size $r$ can remain unallocated if
there is an agent (or agents) with $s$ skill-hour (total) of at least $r$. Since at each allocation epoch
a task should either be allocated totally or not at all (i.e., $x<r$ hours cannot be allocated),
it may happen that some agent hours are wasted, as that does not meet the task allocation 
requirement. 

Note that since any job requirement is less than $\bar{r}=\max_{j,s} r_{j,s}$, at most
$\bar{r}$ agent-skill-hours can be wasted. 

Let $H_s(t)$ be the process of unallocated job-hours for skill $s$ after the allocation at
epoch $t$. Then for all $t$,
\[
H_s(t+1) \le \left|H_s(t) + \sum_j A_j(t) r_{j,s} - \sum_i U_i(t) h_{i,s} + \bar{r}\right|^{+}.
\]
This implies that process $G_s(t)$ given by
$G_s(t+1)=\left|G_s(t) + \sum_j A_j(t) r_{j,s} - \sum_i U_i(t) h_{i,s} + \bar{r}\right|^{+}$ 
bounds $H_s(t)$.

$G_s(t)$ has dynamics of a queue with arrival process $X_s(t) = \sum_j A_j(t) r_{j,s} + \bar{r}$ and
departure process $Y_s(t)=\sum_i U_i(t) h_{i,s}$. Let $\mc{A}_j(\theta)=\EX\left[e^{\theta A_j(t)}\right]$
and $\mc{U}_i(\theta)=\EX\left[e^{\theta U_i(t)}\right]$ for $j \in [N]$ and $i \in [M]$.

For $\theta \in \R$, then,
\begin{align}
\EX\left[e^{\theta (X_s(t)-Y_s(t))}\right]  &= \EX\left[e^{\theta X_s(t)}\right] \EX\left[e^{-\theta Y_s(t)}\right] \nonumber \\
&=\EX\left[e^{\theta \sum_j A_j(t) r_{j,s} + \bar{r}}\right]   
\EX\left[e^{-\theta \sum_i U_i(t) h_{i,s}}\right]   \nonumber \\
&= e^{\theta \bar{r}} \prod_j\EX\left[e^{\theta A_j(t) r_{j,s}}\right]  \prod_i\EX\left[U_i(t) h_{i,s}\right] \label{eq:GA1} \\
&= e^{\theta \bar{r}} \prod_j \mc{A}_j(\theta r_{j,s}) \prod_i \mc{U}_i(-\theta h_{i,s}) \nonumber \\
&= \exp \left(\theta \bar{r} + \sum_j \log \mc{A}_j(\theta r_{j,s}) + \sum_i \log \mc{U}_i(-\theta h_{i,s}) \right)\mbox{.} \nonumber
\end{align}

First consider the Gaussian-dominated case. Since the process variance is no more than the mean and
the moment generating function of the variance is upper-bounded by that of a zero-mean Gaussian:
\begin{align*}
\log \mc{A}_j(\theta r_{j,s}) &\le \lambda_j \theta r_{j,s} + \lambda_j \frac{\left(\theta r_{j,s}\right)^2}{2} \\
\log \mc{U}_i(-\theta h_{i,s}) &\le - \mu_j \theta h_{i,s} + \mu_j \frac{\left(\theta h_{i,s}\right)^2}{2}\mbox{.} 
\end{align*}

Note that for any two functions $k_1 x^2$ and $k_2 x$, $\lim_{x \to 0} k_2 x/k_1 x^2 = \infty$, and
hence for any $\epsilon\in(0,1)$ there exists $x^*>0$ such that for all $x<x^*$, $k_1 x^2/k_2 x <
\epsilon$. Hence for any $\epsilon \in (0,1)$,
there exist $\theta^*_{j,s}, \theta^*_{i,s}>0$, for all $i, j, s$ such that for all 
$\theta < \theta^*=\min_{i,j,s} (\theta^*_{j,s},\theta^*_{i,s}$, 
\begin{align}
\log \mc{A}_j(\theta r_{j,s}) &\le \lambda_j \theta^* r_{j,s} (1+\epsilon) \label{eq:GA2}\\
\log \mc{U}_i(-\theta h_{i,s}) &\le - \mu_i \theta h_{i,s} (1-\epsilon) \label{eq:GA2A}
\end{align}

Note that since $N$, $S$, and $M$ are finite and $\theta^*_{j,s}, \theta^*_{i,s}>0$, 
for all $i, j, s, \theta^*>0$. Moreover, note that $\theta^*$ does not depend on $\vlambda, \bm{\mu}$ since the ratio of 
the linear and quadratic terms in the log moment generating functions are independent 
of $\vlambda$ and $\bm{\mu}$.

As $e^{\theta} - 1 = \sum_{k=1}^\infty \frac{\theta^k}{k!}$, for the Poisson-dominated case
we have
\begin{align*}
\log \mc{A}_j(\theta r_{j,s}) &\le \lambda_j \sum_k \frac{\left(\theta r_{j,s}\right)^k}{k!} \\
\log \mc{U}_i(-\theta h_{i,s}) &\le \mu_j \sum_k \frac{\left(-\theta h_{i,s}\right)^k}{k!} 
\end{align*}

Again, by the same argument, we can have a $\theta^*$ for which (\ref{eq:GA2}) and (\ref{eq:GA2A})
are satisfied. Thus, for all $\theta<\theta^*$ we have:
\begin{align}
\EX\left[e^{\theta (X_s(t)-Y_s(t))}\right] &\le \exp \left(\theta \bar{r} + \sum_j \lambda_j \theta^* r_{j,s} (1+\epsilon) -
\sum_i \mu_i \theta h_{i,s} (1-\epsilon)\right) \nonumber \\
& \le \exp \left(\theta \left(\bar{r} - \sum_i \mu_i h_{i,s}(\alpha - \epsilon)\right)\right) \mbox{.}
\label{eq:GA3} 
\end{align}
Note \eqref{eq:GA3} follows from the fact $\lambda \in (1-\alpha) \C^{out}$. As $\epsilon>0$ can be chosen
arbitrarily small, we can have $\alpha-\epsilon>0$. Since $\sum_i \mu_i h_{i,s}>\sum_j (1-\alpha)\lambda_j r_{j,s} $ and $\sum_j \lambda_j r_{j,s}$ scales with $\lambda(N)$, for sufficiently large $\lambda_{\alpha}$
with $\lambda_j \ge \lambda_{\alpha}$ for all $j$, we have 
$\bar{r} - \sum_i \mu_i h_{i,s}(\alpha - \epsilon) \le - \gamma \sum_i \mu_i h_{i,s}(\alpha - \epsilon)$,
for some $\gamma>0$. Thus, we have for some $\theta>0$,
\begin{equation}
\EX\left[e^{\theta (X_s(t)-Y_s(t))}\right] \le \exp \left(-\theta K\right) \mbox{,}
\end{equation}
where $K$ scales with $\vlambda$.

Now by Loynes' construction: 
\[
G_s(t) = \max_{\tau<t} \sum_{k=\tau}^t (X_s(t)-Y_s(t))\mbox{.}
\]
Hence,
\begin{align}
\PR\left(G_s>g\right) 
&\le \frac{\EX\left[e^{\theta G_s}\right]}{e^{\theta g}} \nonumber \\
& \le e^{-\theta g} \EX\left[e^{\theta \max_\tau \sum_{t=0}^\tau (X_s(t)-Y_s(t))}\right] \nonumber \\
& \le e^{-\theta g} \sum_{\tau=1}^\infty \left(\EX\left[e^{\theta (X_s(t)-Y_s(t))}\right]\right)^\tau \label{eq:GA4} \\
& \le e^{-\theta g} \sum_{\tau=1}^\infty \exp \left(-\theta \tau K\right) \nonumber \\
& \le e^{-\theta g} \frac{1}{1-\exp\left(-\theta K\right)}. \nonumber
\end{align}
This in turn implies
\begin{align}
\PR\left(\max_s G_s>c \log N\right)
& \le S \PR\left(G_s > c \log N\right) \nonumber \\
& \le \frac{S e^{-\theta c \log N}} {1-\exp\left(-\theta K\right)} \nonumber \\
& \le \frac{S}{N^{\theta c}} \frac{1}{1-\exp\left(-\theta K\right)}\mbox{.}\nonumber
\end{align}

Note that the total number of unallocated jobs in the system is upper-bounded by
$(\min_{j,s: r_{j,s}>0} r_{j,s})^{-1}\sum_s G_s$, as any unallocated job has
at least $\min_{j,s: r_{j,s}>0} r_{j,s}$ skill-hours unallocated. Hence the total
number of unallocated tasks in the system, $Q$, satisfies:
\begin{align}
\PR\left(Q > c S \log N\right) 
&\le \PR \left(\sum_s G_s> c \min_{j,s: r_{j,s}>0} r_{j,s} S \log N\right) \nonumber \\
&\le \PR\left(\max_s G_s > c \min_{j,s: r_{j,s}>0} r_{j,s}\log N\right) \frac{S}{N^{\theta c \min_{j,s: r_{j,s}>0} r_{j,s}}} \frac{1}{1-\exp\left(-\theta K\right)}\mbox{.} \nonumber 
\end{align}

Note that $\theta^*>0$ does not depend on $\vlambda$, so $\frac{1}{1-\exp\left(-\theta K\right)}$
is $O(1)$. As $S=O(N)$, we can choose a $c>0$ such that
$\theta^* c \min_{j,s: r_{j,s}>0} r_{j,s} > 3$ and hence,
$\PR\left(Q > c S \log N\right) \le o(N^{-2})$.

\subsection{Proof of Proposition \ref{prop:greedyAgentNDunstable}}
\label{sec:prop:greedyAgentNDunstable}
We prove this proposition by constructing a simple (but general) system and show that the system is not stable via a domination argument with a Markov chain that is not positive recurrent. 

Consider $N=1$, $M=2$, and $S>1$ being even. Let $r=\mbf{1}$, $h_1 = (1,1, \cdots \frac{S}{2} \ \mbox{terms}, 0, 0, \cdots)$
and $h_2 = (0,0, \cdots \frac{S}{2} \ \mbox{terms}, 1, 1, \cdots)$. Let the arrival to the system be i.i.d.\ with
\[
\begin{cases}
(1-\alpha-\delta)\lambda, & \mbox{w.p. } (1-\epsilon) \\
2 \lambda, & \mbox{w.p. } \epsilon, 
\end{cases}
\]
such that $(1+\alpha)\epsilon - \delta(1-\epsilon) = 0$, resulting in mean arrival rate $(1-\lambda)$ and
variance $<\lambda$. One possible construction is to take $\epsilon=\frac{1}{2\lambda}$ and $\delta$ accordingly.
The agent-availability process is considered to be $U_1 = U^2 = \lambda$. Note that both arrival and agent
availability processes are Gaussian-dominated as well as Poisson-dominated.

First, consider the case where greedy picking of tasks by the agents may be adversarial. Each agent
picks all the tasks that it can take from the job and so agents of different types pick from a different half of the $S$ skill-parts. Thus if there are $\ge 2\lambda$ jobs, in worst case (where adversary gives the 
job to the agent) agents of type $1$ may pick $\frac{S}{2}$ job-parts of $\lambda$ jobs, while agents of
type $2$ pick parts of other $\lambda$ jobs. Hence no job is actually allocated at that
allocation epoch. By the next epoch at least $(1-\alpha) \lambda$ jobs have come and if the agents 
pick jobs in an adversarial manner, again no job is actually allocated. Thus the number of unallocated
jobs keep growing after it hits $2\lambda$ once. Note that since there is a positive probability of $\ge 2\lambda$ arrivals, with strictly positive probability, the number of unallocated jobs grows without bound.

Next, we prove the case where greedy picking of the tasks by the agents is random, i.e., an agent picks
all $S/2$ parts of a randomly selected job (without replacement). As arrivals and availability are 
i.i.d., we can describe the number of unallocated jobs by a Markov chain $Q(t)$. 
Note that for $Q\le\lambda$, $\PR(Q\to0)=1$. On the other hand, $0<\PR(0\to Q)<1$ for $Q<2 \lambda$. 

Consider any $Q=n \lambda$ for $n>1$. Note that {{$\PR(Q\to x)=$}} for $x<(n-1) \lambda$ as no more than
$\lambda$ jobs can be scheduled, because there are $\lambda$ agents of each type. Again, 
\[
\PR\left(Q \ \mbox{decreases at least by 1}\right) \le 1 - \left(1-\frac{\lambda}{Q+\lambda-\alpha\lambda}\right)^\lambda
\]
because there are at least $(1-\alpha)\lambda$ arrivals and each type picks $\lambda$ agents 
randomly and this is the probability that the picked sets have a non-empty intersection. Again, as there
are $2\lambda$ arrivals w.p. $\ge \epsilon$ we have
\begin{align}
\PR\left(Q \mbox{ increases by at least } \lambda\right) \ge \epsilon \left(1-\frac{\lambda}{Q+2\lambda}\right)^\lambda. \nonumber
\end{align}

Based on computation of transition probabilities for each transition, it follows that for all $k\ge 1$,
the probability of $Q$ decreasing by at least $1$ as well as the probability of $Q$ decreasing by $k$ 
is decreasing with $Q$, whereas the probability of $Q$ increasing by $k$ increases.

Hence, we can dominate the above chain by another chain $\hat{Q}$ on $\lambda \Z^+$ with transition probabilities
\begin{align}
\PR\left(\lambda n \to \lambda (n+1)\right) &= \epsilon \left(1-\frac{1}{n+1}\right)^\lambda \nonumber \\
\PR\left(\lambda (n+1) \to \lambda n\right) &= 1 - \left(1-\frac{1}{n+1}\right)^\lambda \nonumber.
\end{align}

Now the chain $\hat{Q}(t)$ is a birth-death chain. If it has a finite (summable over states) invariant, then  that is unique. We first assume that the invariant is $\pi$ and then show that it is not summable to prove that
it is not positive recurrent.\footnote{An alternate proof follows from noting that if we take a Lyapunov $\hat{Q}$ itself, then
it has bounded jumps and it is easy to check that after certain $Q>0$ the drift is strictly positive, and invoke the 
Foster-Lyapunov (converse) theorem  for irreducible chain with bounded absolute drift.}
As this is a birth-death chain the invariant measure must satisfy:
\[
\pi(n+1) = \pi(n) \frac{\epsilon \left(1-\frac{1}{n+1}\right)^\lambda}{1 - \left(1-\frac{1}{n+1}\right)^\lambda}. \nonumber
\]
Since
\[
\frac{\epsilon \left(1-\frac{1}{n+1}\right)^\lambda}{1 - \left(1-\frac{1}{n+1}\right)^\lambda} \to \infty\]
as $n \to \infty$ for any finite $\lambda>0$, this shows that $\pi$ is not finite.

\subsection{Proof of Theorem \ref{thm:greedyJob}}
\label{sec:thm:greedyJob}

Consider the different types of unallocated jobs.
These are given by $\{Q_{j}(t): j \in [N]\}$. 

Consider the following processes: for each $s \in [S]$,
$Q^s(t)=\sum_{j:r_{j,s}>0} Q_{j} r_{j,s}$ which represent
the number of unserved hours of skills $s$ over all jobs. 

We now construct another process $\tilde{Q}$ s.t. it dominates
the process $\sum_s Q^s$. So, if we can show upper-bound on $\tilde{Q}$,
then the same bound applies for $\sum_s Q^s$. Hence, in turn we get a
bound for $\{Q_{j}(t)\}$ (as $\min\{r_{j,s}>0\}=\Theta(1)$ by the
assumption that $\{r_{j,s}\}$ do not scale with the system size).

Towards constructing a suitable $\tilde{Q}$ we make the following observation
about the dynamics of $Q^s$ and $\{Q_{j}\}$. 
At each time $t$,
$\sum_j A_{j,1}(t) r_{j,s}$ amount of $s$ skill hour is brought to add
to $Q^s$. Also, this queue gets some service depending on the available
agent hours.

At time $t$, $\sum_m U_m(t) h_{m,s}$ $s$-skill hour of service is brought
by the agents. 

For a job to be allocated, all tasks of it must find an allocation.
Hence, for a job in type $j$-job to find an allocation it must get $r_{j,s}$
hours of service from each skill $s$. Thus at any time $t$ any skill $s$ queue
gets a service of at least

$$\min_{s \in [S]} \sum_m U_m h_{m,s} - \bar{r},$$

where $\bar{r}=\max\{r_{j,k,s}\}$. This is because of the following. 
For each skill $\sum_s U_m h_{m,s}$ hour is available. Note that a
job can be allocated if all its tasks find allocations, coverse of which is also
true. That is if all tasks of a step found allocation then the step can be 
allocated. As $\min_{s \in [S]} \sum_s U_m h_{m,s}$
hours of service is brought by the agents for each skill, at 
least $\min_{s \in [S]} \sum_s U_m h_{m,s} - \bar{r}$ of $s$-skill
hours are served (because a maximum of $\bar{r}$ can be wasted, as no
task is of size more than $\bar{r}$). 

Also, note that the amount of required service brought to the 
queue $Q^s$ at time $t$ is upper-bounded by

$$\max_{s \in [S]} \sum_j A_{j}(t) r_{j,s}$$

Consider a process $\tilde{Q}^s$ with evolution
\begin{align}
\tilde{Q}^s(t+1) & = \max(\tilde{Q}^s(t) +  \max_{s \in [S]} \sum_j A_{j}(t) r_{j,s} \nonumber \\
& \ - \min_{s \in [S]} \sum_m U_m h_{m,s} + \bar{r}, 0). \nonumber 
\end{align}
Note that given $\tilde{Q}^s(t_0)\ge Q^s(t_0)$ at some $t_0$, the same holds true
for all $t \ge t_0$. This is because for $x, a, b \ge 0$ and $x', a', b' \ge 0$, with
$x\ge x'$, $a\ge a'$ and $b\le b'$
$$\max(x+a-b,0) \ge \max(x'+a'-b',0),$$
and hence, the monotonicity propagates over time.

Thus, to bound $\sum_s Q^s$ it is sufficient to bound $\sum_s \tilde{Q}^s(t)$.
Note that each of $\tilde{Q}^s$ has exactly same evolution, so let us
consider 
$$\tilde{Q}:=S\tilde{Q}^1,$$ 
which bounds $\sum_s Q^s$.

From the evolution:
\[
\tilde{Q}(t+1) = \max(\tilde{Q}(t) +  S \max_{s \in [S]} \sum_j A_{j}(t) r_{j,s} - S \min_{s \in [S]} \sum_m U_m h_{m,s} + \bar{r}, 0)
\]
we can write the Loynes' construction for this process which has the same distribution as
this process (and for simplicity we use the same notation, as we are interested in 
the distribution).
\begin{equation}
\tilde{Q}^1(0) = \max_{\tau\le 0} \sum_{\tau\le t \le 0}(S \max_{s \in [S]} \sum_j A_{j}(t) r_{j,s} - S \min_{s \in [S]} \sum_m U_m h_{m,s} + \bar{r}) \mbox{,}
\end{equation}
assuming that the process started at $-\infty$.

Let us define $X_s(t)$ and $Y_s(t)$ as follows:
$X_s(t):=\sum_j A_{j}(t) r_{j,s}$ and 
$Y_s(t):=\sum_m U_m h_{m,s}$.
Then, 
$$\tilde{Q}^1(0) = \max_{\tau\le 0}
\sum_{\tau\le t \le 0}S(\max_{s} X_s(t) - \min_s Y_s(t) + \bar{r}).$$

Now, for any $\theta>0$
\begin{align}
\PR(\sum_{j} Q_{j,1} > \bar{r} q) 
&\le \PR(\sum_s Q^s_1 > q) \nonumber \\
&\le \PR(\tilde{Q}_1(0) > q)  \nonumber \\
&= \PR( \theta \tilde{Q}_1(0) > \theta q) \label{eq:prioGreedy1} \\
&= \PR(\exp(\theta \tilde{Q}_1(0)) > \exp(\theta q)) \nonumber \\
&\le \EX[\exp(-\theta q)] \EX[\exp(\theta \tilde{Q}_1(0))]\mbox{.} \nonumber
\end{align}
Now,
\begin{align}
\EX[\exp(\theta \tilde{Q}_1(0))] 
&= \EX[\exp(\theta S \left(\max_{\tau\le 0}
\sum_{\tau\le t \le 0}(\max_{s} X_s(t) - \min_s Y_s(t) + \bar{r})\right))] \nonumber \\
&\le\sum_{\tau\le 0} \EX[\exp(\theta S 
\sum_{\tau\le t \le 0}(\max_{s}X_s(t) - \min_s Y_s(t) + \bar{r}))],\label{eq:prioGreedy2}
\end{align}
where the inequality in \eqref{eq:prioGreedy2} 
follows because for any random variables $\{Z_j\}$,
$\exp(\theta \Z_j)$ are positive random variables and sum of positives are
more than their maximum.

Next, we bound the term within the summation over $\tau\le 0$ in \eqref{eq:prioGreedy2}.
\begin{equation}
\EX[\exp(\theta S \sum_{\tau\le t \le 0}(\max_{s} X_s(t) - \min_s Y_s(t) + \bar{r}))] \le \prod_{\tau \le t \le 0} \EX[\exp(\theta (\max_{s}X_s(t) - \min_s Y_s(t) + \bar{r})))]\mbox{.}
\label{eq:prioGreedy3}
\end{equation}
Inequality in \eqref{eq:prioGreedy3} follows because $X_s(t)$, $Y_s(t)$ are i.i.d.\ over
time.

Next we bound the term within the product $\prod_{\tau\le t \le 0}$
in \eqref{eq:prioGreedy3},
\begin{equation}
\EX\left[e^{\theta S  \left(\max_s X_s(t) - \min_s Y_s(t) + \bar{r}\right)}\right] 
\le \sum_{s,s'} \EX \left[e^{\theta S \left(X_s(t) - Y_{s'}(t) + \bar{r}\right)}\right] 
\label{eq:prioGreedy4} 
\end{equation}
Inequality \eqref{eq:prioGreedy4} is due to the same reason as \eqref{eq:prioGreedy2}.

Let $\mc{A}_j(\theta)=\EX\left[e^{\theta A_j(t)}\right]$
and $\mc{U}_m(\theta)=\EX\left[e^{\theta U_m(t)}\right]$ for $j \in [N]$ and $m \in [M]$.
For $\theta \in \R$, then,
\begin{align}
\EX\left[e^{\theta (X_s(t)-Y_{s'}(t))}\right] 
&= \EX\left[e^{\theta X_s(t)}\right] \EX\left[e^{-\theta Y_{s'}(t)}\right] \nonumber \\
&=\EX\left[e^{\theta \sum_j A_j(t) r_{j,s} + \bar{r}}\right]   
\EX\left[e^{-\theta \sum_i U_i(t) h_{i,s}}\right]   \nonumber \\
&= e^{\theta \bar{r}} \prod_j\EX\left[e^{\theta A_j(t) r_{j,s}}\right]  \prod_i\EX\left[e^{-\theta U_i(t) h_{i,s'}}\right] \label{eq:GA1} \\
&= e^{\theta \bar{r}} \prod_j \mc{A}_j(\theta r_{j,s}) \prod_i \mc{U}_i(-\theta h_{i,s'}) \nonumber \\
&= \exp \left(\theta \bar{r} + \sum_j \log \mc{A}_j(\theta r_{j,s}) + \sum_i \log \mc{U}_i(-\theta h_{i,s'}) \right)\mbox{.} \nonumber
\end{align}

Note that as $\vlambda \in \alpha \mc{C}$, by the definition of $\mc{C}^O$,
$\sum_j \lambda_j r_{j,s} <\alpha \sum_m \mu_m h_{m,s}$ and by assumption
$|\sum_m \mu_m h_{m,s} - \sum_m \mu_m h_{m,s'}|\le \mbox{subpoly}(N)$ which is used
in the following.

First consider the Gaussian-dominated case. Since the process variance is no more than mean and
the moment generating function of the variance is upper-bounded by that of a zero-mean Gaussian:
\begin{align*}
\log \mc{A}_j(\theta r_{j,s}) &\le \lambda_j \theta r_{j,s} + \lambda_j \frac{\left(\theta r_{j,1,s}\right)^2}{2} \\
\log \mc{U}_i(-\theta h_{i,s}) &\le - \mu_j \theta h_{i,s} + \mu_j \frac{\left(\theta h_{i,s}\right)^2}{2}\mbox{.} 
\end{align*}

Note that for any two functions $k_1 x^2$ and $k_2 x$, $\lim_{x \to 0} k_2 x/k_1 x^2 = \infty$, and
hence for any $\epsilon\in(0,1)$ there exists $x^*>0$ such that for all $x<x^*$, $k_1 x^2/k_2 x <
\epsilon$. Hence for any $\epsilon \in (0,1)$,
there exist $\theta^*_{j,s}, \theta^*_{i,s}>0$, for all $i, j, s$ such that for all 
$\theta < \theta^*=\min_{i,j,s} (\theta^*_{j,s},\theta^*_{i,s}$, 
\begin{align}
\log \mc{A}_j(\theta r_{j,1,s}) &\le \lambda_j \theta^* r_{j,s} (1+\epsilon) \label{eq:GA2}\\
\log \mc{U}_i(-\theta h_{i,s}) &\le - \mu_i \theta h_{i,s} (1-\epsilon) \label{eq:GA2A}
\end{align}

Note that since $N$, $S$, and $M$ are finite and $\theta^*_{j,s}, \theta^*_{i,s}>0$, 
for all $i, j, s, \theta^*>0$. Moreover, note that $\theta^*$ does not depend on $\vlambda, \bm{\mu}$ since the ratio of 
the linear and quadratic terms in the log moment generating functions are independent 
of $\vlambda$ and $\bm{\mu}$.

As $e^{\theta} - 1 = \sum_{k=1}^\infty \frac{\theta^k}{k!}$, for the Poisson-dominated case
we have
\begin{align*}
\log \mc{A}_j(\theta r_{j,s}) &\le \lambda_j \sum_k \frac{\left(\theta r_{j,s}\right)^k}{k!} \\
\log \mc{U}_i(-\theta h_{i,s}) &\le \mu_j \sum_k \frac{\left(-\theta h_{i,s}\right)^k}{k!} 
\end{align*}

Again, by the same argument, we can have a $\theta^*$ for which (\ref{eq:GA2}) and (\ref{eq:GA2A})
are satisfied. As $|\sum_i \mu_i h_{i,s} - \sum_i \mu_i h_{i,s'}|=o(N^{\delta})$, for all $\delta>0$,
and $\sum_i \mu_i h_{i,s}=\Omega(N^c), c>0$, for all $\theta<\theta^*$ we have:
\begin{align}
\EX\left[e^{\theta (X_s(t)-Y_{s'}(t))}\right] 
&\le \exp \left(\theta^* \bar{r} + \sum_j \lambda_j \theta^* r_{j,s} (1+\epsilon) -
\sum_i \mu_i \theta h_{i,s} (1-\epsilon) + \theta^* o(\mu_i \theta h_{i,s})\right) \nonumber \\
& \le \exp \left(\theta \left(\bar{r} - \sum_i \mu_i h_{i,s}(\alpha - 2 \epsilon)\right)\right) \mbox{.}
\label{eq:GA3} 
\end{align}
Note \eqref{eq:GA3} follows from the fact $\lambda \in (1-\alpha) \C^{out}$. As $\epsilon>0$ can be chosen
arbitrarily small, we can have $\alpha-2\epsilon>0$. Since $\sum_i \mu_i h_{i,s}>\sum_j (1-\alpha)\lambda_j r_{j,s} $ and $\sum_j \lambda_j r_{j,1,s}$ scales with $\lambda(N)$, for sufficiently large $\lambda_{\alpha}$
with $\lambda_j \ge \lambda_{\alpha}$ for all $j$, we have 
$\bar{r} - \sum_i \mu_i h_{i,s}(\alpha - \epsilon) \le - \gamma \sum_i \mu_i h_{i,s}(\alpha - \epsilon)$,
for some $\gamma>0$. Thus, we have for some $\theta>0$,
\begin{equation}
\EX\left[e^{\theta S (X_s(t)-Y_{s'}(t))}\right] \le \exp \left(-\theta S K(N) \right) \mbox{,}
\end{equation}
where $K(N)$ scales with $N$ no slower than 
$\sum_{s:r_{j,1,s}>0} \lambda_j(N) =\Omega(N^c)$, $c>0$.

Thus,
\[
\EX\left[e^{\theta S (\max_s X_s(t) - \min_s Y_s(t))}\right] \le S^2 \exp \left(-\theta S K(N) \right) \mbox{.}
\]
Hence, from \eqref{eq:prioGreedy2}, \eqref{eq:prioGreedy3}, and \eqref{eq:prioGreedy4}
we have that
\begin{align}
\EX[\exp(\theta \sum_s \tilde{Q}^s(0))]
&= \EX[\exp(\theta S \tilde{Q}^1(0))] \nonumber \\
&= \EX[\exp(\theta \tilde{Q}(0))] \nonumber \\
&\le \sum_{\tau \le 0}  S^{|2\tau|}
\exp(-\theta S K(N) |\tau|) \nonumber \\
&\le c', \nonumber
\end{align}
because $S^2 < \exp(\theta^* S K(N))$ for all sufficiently large $N$.

Note that though we proved $\EX[\exp(\theta {Q}(t))]<c'$ for $t=0$, this holds for
any finite $t$ with exactly the same proof. 
Hence, 
\[
\PR(Q> c \log N) \le \PR(\tilde{Q} > c \log N) \le c' e^{-c \theta^* \log N}\mbox{,}
\]
which gives the result for an appropriate choice of $c$.

\subsection{Proof of Theorem \ref{thm:altStability}}
\label{sec:thm:altStability}
To prove that any $\vlambda \in \C^L$ is stabilizable it is sufficient
to invoke Thm.~\ref{thm:JSQandSRMW} whose proof is below.
To show that $\vlambda^O \notin \C^I$ is not stabilizable, we take an
approach similar to the proof of Thm.~\ref{thm:capacityGen}.

Note that the set $\{\vlambda: \vlambda=\sum_l \vlambda^{l}, 
\vlambda^l \in \C^l_{\Gamma}\}$ is convex. Also, if 
$\vlambda'\le\vlambda$ (component-wise) for some $\vlambda$ 
belonging to the set, then $\vlambda' \in \C^I$. That is,
$\C^I$ is coordinate convex.
Hence for any $\vlambda \notin \mbox{closure of } \C^L$, there exists
a hyperplane $h\ge 0$ that strictly separates it from $\C^L$, i.e.,
for any $\{\vlambda^l \in \C^l\}$, for some $\epsilon>0$
\[
h^T \lambda^O \ge h^T \sum_l \vlambda^l + \epsilon 
\]
Note that at any epoch $t$, number of $j$ jobs allocated $\Delta_j(t)=\sum_l\Delta^l_j(t)$,
where $\Delta^l_j(t)$ is the number of $j$ jobs allocated to category $l$ agents.
Following similar steps as in the proof of Thm.~\ref{thm:capacityGen}, the result follows.

\subsection{Proof of Theorem \ref{thm:JSQandSRMW}}
\label{sec:thm:JSQandSRMW}
Let $Q_{l,j,s}(t)$ be the number of unallocated $(j,s)$ tasks in
the $l$th pool and $A_{j}(t)$ be the number of arrived jobs of type
$j$. Let $A_{l,j}(t)$ be the number of jobs that are sent to pool $l$
and $\sum_l A_{l,j} = A_j(t)$. At any pool $l$ MaxWeight is followed and
$\hat{D}_{l,j,s}(t)$ is the number of allocated $(j,s)$ tasks in
pool $l$.

Consider a Lyapunov function $L(\mbf{Q})=\sum_{l,j,s} Q_{l,j,s}^2$. Then:
\begin{align}
&\EX\left[L\left(\mbf{Q}(t+1)\right) - L\left(\mbf{Q}(t)\right)|\mbf{Q}(t)\right] \nonumber \\
&\quad\le \EX\left[\sum_{l,j,s}  Q_{l,j,s}(t) A_{l,j}(t)|\mbf{Q}(t)\right]  - \EX\left[\sum_{l,j,s} Q_{l,j,s}(t) \hat{\Delta}_{l,j,s}(t)|\mbf{Q}(t)\right]  + \EX\left[\sum_{l,j,s} \left(A^2_{l,j}(t)+\hat{\Delta}^2_{l,j,s}\right)|\mbf{Q}(t)\right] \label{eq:JSQ1}
\end{align}
The last term can be bounded by noting: 
\begin{align}
\sum_{l,j,s} \left(A^2_{l,j}(t)+\hat{\Delta}^2_{l,j,s}\right) &\le \left(\sum_{l,j,s} \left(A_{l,j}(t) + \hat{\Delta}_{l,j,s}\right)\right)^2 \nonumber \\
& = \left(\sum_{j,s} A_{j}(t) + \sum_{l,j,s} \hat{\Delta}_{l,j,s}\right)^2 \nonumber \\
& \le \left(\sum_{j,s} A_{j}(t) + \frac{1}{\min(r_{j,s}>0)} \sum_{l,i,s} h^l_{i,s} U^l_{i,s}\right)^2 \label{eq:JSQ2} \\
& \le B' < \infty. \label{eq:JSQ3}
\end{align}
Note \eqref{eq:JSQ2} follows similarly as \eqref{eq:SRMW4}, whereas \eqref{eq:JSQ3} follows because the arrival
and agent-availability processes have bounded second moments.

To bound the first term:
\begin{align}
\EX\left[\sum_{l,j,s} Q_{l,j,s}(t) A_{l,j}(t)|\mbf{Q}(t)\right] &= \EX\left[\sum_{l,j} \left(\sum_s Q_{l,j,s}(t)\right) A_{l,j}(t)|\mbf{Q}(t)\right] \nonumber \\
& \le \EX\left[\sum_{l,j,s} \frac{\lambda^l_j}{\lambda_j} A_j(t)  Q_{l,j,s}(t)|\mbf{Q}(t)\right]
\label{eq:JSQ4} \\
& \le \sum_l \left(\sum_{j,s} Q_{l,j,s}(t) \lambda^l_j\right) \label{eq:JSQ5}
\end{align}
where \eqref{eq:JSQ4} is because of the fact JLTT-MWTA sends all arrivals of type $j$ to 
the pool $l$ with minimum $\sum_s Q_{l,j,s}(t)$.

On the other hand,
\begin{equation}
\EX\left[\sum_{l,j,s} Q_{l,j,s}(t) \hat{\Delta}_{l,j,s}(t)|\mbf{Q}(t)\right] = \sum_l \EX\left[\sum_{j,s}  Q_{l,j,s}(t) \hat{\Delta}_{l,j,s}(t)|\mbf{Q}^l(t)\right], \label{eq:JSQ6}
\end{equation}
because at epoch $t$ each pool $l$ runs MaxWeight based on only $\mbf{Q}_{l}(t)$ and
$\{\hat{\Delta}_{l,j,s}:j,s\}$ is independent of $\{A_{l',j}(t), Q_{l',j,s}(t):l'\neq l\}$
given $\mbf{Q}_{l}(t)$.

For every $l$, we can compute the difference between the $l$th term of \eqref{eq:JSQ5} and that of
\eqref{eq:JSQ6}, which is similar to the first term of \eqref{eq:SRMW4}. If 
$\vlambda+\epsilon \in \C^L$, then $\vlambda^l+\frac{\epsilon}{L} \in \C^l$, hence following the same
steps as in the proof of Thm.~\ref{thm:SRMW} we have
\[
\left(\sum_{j,s}  Q_{l,j,s}(t) \lambda^l_j\right)-\EX\left[\sum_{j,s} Q_{l,j,s}(t) \hat{\Delta}_{l,j,s}(t)|\mbf{Q}^l(t)\right] \le -\frac{\epsilon}{L} \sum_{j,s}  Q_{l,j,s}(t) \mbox{.}
\]

This in turn implies
\[
\EX\left[L\left(\mbf{Q}(t+1)\right) - L\left(\mbf{Q}(t)\right)|\mbf{Q}(t)\right] \le B' - \frac{\epsilon}{L} \sum_{l,j,s} Q_{l,j,s}(t).
\]

Following similar steps as in proof of Thm.~\ref{thm:SRMW}, we obtain 
$\lim \sup_{t \to \infty} \sum_{l,j,s} \EX\left[Q_{l,j,s}(t)\right] < \infty$,
which in turn implies that for all $j,s,l$, 
$\lim \sup_{t \to \infty} \EX\left[Q_{l,j,s}(t)\right] < \infty$. This proves the theorem
since for all $j$, $Q_j(t) \le \sum_{l,s} Q_{l,j,s}(t)$.

\subsection{Proof of Theorem \ref{thm:altOuter}}
\label{sec:thm:altOuter}
It is sufficient to prove that $\C^I \subseteq \C^O$.
Consider any $\vlambda \in \C^I$, then by definition of $\C^I$,
$\vlambda = \sum_l \vlambda^l$, where for all $l$, $\vlambda^l \in \C^l$.
By the characterization of the outer region of single category systems
$\C^l \subseteq \C^{out}_l$, $\lambda^l \in \C^{out}_l$. 
Thus $\vlambda = \sum_l \vlambda^l$ for $\vlambda^l \in \C^{out}_l$, 
for all $l$. Thus by definition of $\C^O$,
$\vlambda \in \C^O$, which completes the proof.

\subsection{Proof of Theorem \ref{thm:ImprJSQSRMW}}
\label{sec:thm:ImprJSQSRMW}
We find a high-probability bound on the
number of unallocated jobs of type $j$ and then bound the maximum number of jobs across type. 
To bound the number of unallocated jobs, we use stochastic domination based on
the nature of the method of splitting job arrivals across different pools.

Consider the dynamics of $Q^*_j(t)=\max_l Q_{l,j}(t)$. Let $A_{j,l}(t)$ be the
number of jobs of type $j$ that were directed to pool $l$.
In the Improvised JSQ step, jobs are sent one-by-one with minimum backlog and
hence, the queue $l^*$ with $Q^{l^*}_j(t)=Q^*_j(t)$ gets the minimum number of jobs.
Since minimum is less than average, $A_{j,l^*}(t) \le \frac{A_j(t)}{L}$. 
On the other hand,
just before allocation at epoch $t+1$, the total number of $j$-jobs in $l^*$ cannot be
less than the number of jobs in any other $l$ by more than $1$. This is because of the
Improvised JSQ which allocates jobs one-by-one to the lowest backlogged ($N^l_j$) 
queue at that time. Hence, we have
\[
Q^*_j(t) + \lceil \tfrac{A_j(t)}{L} \rceil + 1 \ge Q_{j,l}(t) + A_{j,l}(t)\mbox{.} 
\]

Given $Q_{l,j}$,  for GreedyJob the number of jobs that can be allocated (assuming number of queued 
jobs to be infinite) $\Delta_{l,j}$ is monotonic in $\mbf{U}^l$, i.e., if 
$\mbf{U}^l \ge \mbf{U'}^l$ (component-wise) $\Delta_{l,j} \ge \Delta'_{l,j}$ for all $j$. 
This property will be useful below.

Consider the following dynamics $Q^j(t), j \in [N]$. Arrivals  for each $j$ are according
to $\lceil \frac{A_j(t)}{L} \rceil + 1$ and agent-availability is according to
$U^i(t)=\min_l U^l_i(t)$. This is a single-category system and allocations in this system are
according to GreedyJob. As $U^i(t) \le U^l_i(t)$ for all $i,l$, the number of
allocations (assuming queues to be infinite) satisfies $\Delta^j(t) \le \Delta_{l,j}(t)$. This implies that
for each type $j$ $Q^j(t)$ dominates $Q^*_j(t)$ as the first queue at any epoch has more number of jobs
to be allocated and less number of possible allocations (as $\Delta^j$ is smaller). Thus, it is sufficient to
bound $Q^j(t)$. 

For this we proceed along the lines of the proof of Thm.~\ref{thm:greedyJob}, 
replacing arrivals by $A^j(t)=\lceil \frac{A_j(t)}{L} \rceil + 1$ and agent-availability
by $U^i(t)=\min_l U^l_i(t)$.

Hence, for each skill $s$, the queue of unallocated skill-hours $H^s(t)$ is the arrival
$X^s(t)=\sum_j A^j(t) r_{j,s}$ and the possible amount that can be drained is
$Y^s(t)=\sum_i U^i(t) h_{i,s}$. Hence,
\[
Q^s(t+1) = \left|Q^s(t) + X^s(t) - Y^s(t) + \bar{r}\right|^{+}\mbox{.}
\]

We can follow similar steps by noting that for $\theta>0$:
\begin{align}
\EX\left[e^{-\theta Y^s}\right] 
& = \EX \left[e^{-\theta \sum_i \min_l U^l_i(t) h_{i,s}}\right] \nonumber \\
& = \EX \left[e^{-\theta \min_l \sum_i U^l_i(t) h_{i,s}}\right] \label{eq:ImprJSQ1} \\
& = \EX \left[ \max_l e^{-\theta  \sum_i U^l_i(t) h_{i,s}}\right] \nonumber \\
& \le \sum_l \EX \left[e^{-\theta \sum_i U^l_i(t) h_{i,s}}\right] \nonumber.
\end{align}

On the other hand,
\[
\EX \left[ e^{\theta X^s}\right] \le e^{\theta (\frac{1}{L}+1) \bar{r}} \prod_j \EX \left[e^{\frac{\theta r_{j,s}}{L} A_j(t)}\right]\mbox{.}
\]

Making the assumption on agent arrival rates,
\begin{equation}
\sum_l \EX \left[e^{-\theta \sum_i U^l_i(t) h_{i,s}}\right] \le e^{\log L+O\left(\mbox{subpoly}(N)\right)} \EX \left[e^{-\theta \sum_i U^1_i(t) h_{i,s}}\right]
\end{equation}
Since $\max_{l.l'} |\mu^l_i-\mu^{l'}_i|=\mbox{subpoly}(N)$  for all $i$, $\vlambda \in (1-\alpha) \C^O$ 
implies that $\frac{\vlambda}{L} - \mbox{subpoly}(N) \in \C^{out}$. Thus,  we can use the same steps 
as we did for single-category systems.

For Gaussian-dominated as well as Poisson-dominated cases, following similar steps we can obtain that
for a $\theta^*>0$ (independent of $\vlambda$) and sufficiently large $\vlambda$, 
\[
\EX \left[e^{\theta \left(X^s-Y^s\right)}\right]\le e^{-\theta^* K(\vlambda) + \log L + O\left(\mbox{subpoly}(N)\right)}
\]

Note that $K(\vlambda)$ increases as $\Omega(\min_i\lambda_i)=\Omega(N^c)$ for some $c>0$,
$L=O(1)$, hence there exists 
$\vlambda$ sufficiently large such that 
$-\theta^* K(\vlambda) + \log L + O\left(\mbox{subpoly}(N)\right)$ is strictly negative
and  $\EX \left[e^{\theta \left(X^s-Y^s\right)}\right]<1$. The rest follows similarly as the proof
of Thm.~\ref{thm:greedyAgent}.

\subsection{Proof of Theorem \ref{thm:Drop}}
\label{sec:thm:Drop}
We first consider FD, FND, and IND systems.

Let $\tilde{A}_j(t)$ be the number of accepted jobs of type $j$ between starts of epochs $t-1$ and $t$.
Let $\hat{\Delta}_{j,s}(t)$ be the number of allocated $(j,s)$ tasks by the MaxWeight
part of MWTA before the execution of Task Allocation.
Let $\hat{D}_{j,s}(t)$ be the number of allocated $(j,s)$ tasks at allocation epoch $t$ at the
end of Task Allocation. Then:
\begin{align}
\sum_{j,s:r_{j,s}>0} \left(\tilde{Q}^2_{j,s}(t+1) - \tilde{Q}^2_{j,s}(t)\right)
&= \sum_{j,s:r_{j,s}>0} \left((\tilde{Q}_{j,s}(t) + \tilde{A}_j(t) - \hat{D}_{j,s}(t))^2 - \tilde{Q}^2_{j,s}(t)\right)\nonumber \\
&\le \sum_{j,s:r_{j,s}>0} \left((\tilde{Q}_{j,s}(t) + \tilde{A}_j(t) - \hat{\Delta}_{j,s}(t))^2 - \tilde{Q}^2_{j,s}(t)\right) \nonumber \\
&= \sum_{j,s:r_{j,s}>0} Q_{j,s}(t) \left(\tilde{A}_j(t) - \hat{\Delta}_{j,s}(t)\right) + \sum_{j,s:r_{j,s}>0} \left(\tilde{A}^2_j(t) + \Delta^2_{j,s}(t) \right) \nonumber
\end{align}

Expectation of the second summation (conditioned on $\mbf{Q}(t)$)
can be bounded by noting that $\tilde{A}^2_j(t) \le A_j^2(t)$ and the fact that
the arrival processes have bounded second moment.  
The value $\EX\left[\sum_{j,s:r_{j,s}>0} \Delta^2_{j,s}(t)\right]$ can be bounded similarly as in the
proof of Thm.~\ref{thm:SRMW}. Hence, we consider the expectation of the
second term to be bounded by $B$ independent of $\mbf{Q}$.

\begin{align}
&\EX\left[\sum_{j,s:r_{j,s}>0} \left(\tilde{Q}^2_{j,s}(t+1) - \tilde{Q}^2_{j,s}(t)\right)|\mbf{Q}(t)\right] \nonumber \\
&\le B + \EX\left[\sum_{j,s:r_{j,s}>0} Q_{j,s}(t) \left(\tilde{A}_j(t) - \hat{\Delta}_{j,s}(t)\right)|\mbf{Q}(t)\right] \nonumber \\
&\le B + \EX\left[\EX\left[\sum_{j,s:r_{j,s}>0} Q_{j,s}(t)\left((1-\beta(t)) A_j(t)- \hat{\Delta}_{j,s}(t)\right)  |\mbf{Q}(t), \mbf{A}(t)\right]\mbf{Q}(t)\right]  \nonumber \\
&\le B + \EX\left[\EX\left[\frac{\beta(t)\sum_j A_j(t)}{\nu} + \sum_{j,s:r_{j,s}>0} Q_{j,s}(t)\left((1-\beta(t)) A_j(t) - \hat{\Delta}_{j,s}(t)\right)- \frac{\beta(t)\sum_j A_j(t)}{\nu}|\mbf{Q}(t), \mbf{A}(t) \right]|\mbf{Q}(t)\right]\nonumber \\
&\le B + \EX\left[\EX\left[\frac{\beta(t)\sum_j A_j(t)}{\nu}+ \sum_{j,s:r_{j,s}>0} Q_{j,s}(t)\left((1-\beta(t)) \EX[A_j(t)] - \hat{\Delta}_{j,s}(t)\right)- \frac{\beta(t)\sum_j A_j(t)}{\nu}|\mbf{Q}(t), \mbf{A}(t)\right]\mbf{Q}(t)\right]\nonumber \\
&\le B + \EX\left[\EX\left[\frac{\beta(t)\sum_j A_j(t)}{\nu} + \sum_{j,s:r_{j,s}>0} Q_{j,s}(t)\left((1-\beta(t)) A_j(t) - \hat{\Delta}_{j,s}(t)\right)- \frac{\beta(t)\sum_j A_j(t)}{\nu}|\mbf{Q}(t), \mbf{A}(t)\right]|\mbf{Q}(t)\right] \label{eq:Drop1} \\
&\le B + \EX\left[\EX\left[\frac{\beta^*\sum_j A_j(t)}{\nu} + \sum_{j,s:r_{j,s}>0} Q_{j,s}(t)\left((1-\beta^*) A_j(t) - \hat{\Delta}_{j,s}(t)\right)- \frac{\beta(t)\sum_j A_j(t)}{\nu}|\mbf{Q}(t), \mbf{A}(t)\right]|\mbf{Q}(t)\right] \label{eq:Drop2} \\
&\le B + \frac{1}{\nu}\EX[\left(\beta^* - \beta(t)\right)\sum_j A_j(t)]  + \EX\left[\sum_{j,s:r_{j,s}>0} Q_{j,s}(t)\left((1-\beta^*) A_j(t) - \hat{\Delta}_{j,s}(t)\right)|\mbf{Q}(t)\right] \nonumber
\end{align}
\begin{align}
&\le B + \frac{1}{\nu} \EX[\left(\beta^* - \beta(t)\right)\sum_j A_j(t)]   + \EX\left[\sum_{j,s:r_{j,s}>0} Q_{j,s}(t)\left((1-\beta^*)\lambda_j(t) - \hat{\Delta}_{j,s}(t)\right)|\mbf{Q}(t)\right] \notag
\end{align}
where \eqref{eq:Drop2} follows as $\beta(t)$ minimizes
$\beta \sum_j A_j(t) - \nu \beta  \sum_{j,s: r_{j,s}>0} \tilde{Q}_{j,s}(t) A_j(t)$.

As $(1-\beta^*) \vlambda + \epsilon \mbf{1} \in \C$, following the same steps as in the proof of 
the Thm.~\ref{thm:SRMW}:
\[
\EX\left[\sum_{j,s:r_{j,s}>0} Q_{j,s}(t)\left((1-\beta^*)\lambda_j(t) - \hat{\Delta}_{j,s}(t)\right)|\mbf{Q}(t)\right] \le -\epsilon \sum_{j,s:r_{j,s}>0} Q_{j,s}(t) 
\]
As $\beta^*-\beta(t) \le 1$, we have
\[
\EX\left[\sum_{j,s:r_{j,s}>0} \left(\tilde{Q}^2_{j,s}(t+1) - \tilde{Q}^2_{j,s}(t)\right)\right] 
\le B + \frac{1}{\nu} \sum_j \lambda_j -\epsilon \EX\left[\sum_{j,s:r_{j,s}>0} Q_{j,s}(t)\right]. 
\]

Following again the same steps we show that 
$\lim \sup_{t \to \infty} \EX\left[\sum_{j,s:r_{j,s}>0} Q_{j,s}(t)\right]< \infty$ 
which implies the crowd system is stable.
Hence, we can assume that $\EX\left[\sum_{j,s:r_{j,s}>0} Q_{j,s}(t)\right]<C$ for some $C<\infty$.

Thus we can write 
\[
\EX\left[\sum_{j,s:r_{j,s}>0} \left(\tilde{Q}^2_{j,s}(T) - \tilde{Q}^2_{j,s}(0)\right)\right] \le BT + \frac{1}{\nu} \sum_{t=1}^T \EX[\left(\beta^* - \beta(t)\right)\sum_j A_j(t)], 
\]
which in turn implies
\[
(1-\beta^*) \sum_{j} \lambda_j - \frac{1}{T} \sum_{t=1}^T (1-\beta(t))\sum_j A_j(t) \le \nu B + \frac{\nu}{T} \sum_{j,s:r_{j,s}>0} \tilde{Q}^2_{j,s}(0) 
\]
Since $B$ is a constant depending on arrival and availability statistics, $\nu$ and $T$ can be chosen
to be small and large respectively to ensure that the left side is arbitrarily small. 
The desired result follows by noting that 
\[
\EX\left[\sum_{t=1}^T (1-\beta(t))\sum_j A_j(t)\right] = \EX\left[\sum_{t=1}^T \sum_j \tilde{A}_j(t) \right]\mbox{.}
\]
Also note that since the only requirement is the independence of $\mbf{A}_(t)$ and $\mbf{U}(t)$ across
time, the proof directly extends to settings with non-stationary
arrival and availability processes.

Now consider the ID setting and use the Lyapunov function $\sum_{l,j,s} Q_{l,j,s}^2$. Similar to before,
the Lyapunov drift can be bounded by
\[
B + \EX\left[\sum_{l,j,s:r_{j,s}>0} Q_{l,j,s}(t) \left(\tilde{A}_{l,j}(t) - \hat{\Delta}_{l,j,s}(t)\right)|\mbf{Q}(t)\right]\mbox{.}
\]
Note that $\sum_{l,j,s:r_{j,s}>0} Q_{l,j,s}(t) \tilde{A}_{l,j}(t)$ is equal to
\[
\sum_{l,j,s:r_{j,s}>0} \min_l \left(\sum_s Q_{l,j,s}(t)\right) \tilde{A}_{j}(t)\mbox{,}
\]
as $\tilde{A}_{j}(t)=\sum_l \tilde{A}_{l,j}(t)$ and JLTT ensures that
the jobs are sent to the category with $\min_l \left(\sum_s Q_{l,j,s}(t)\right)$. 
So we have
\begin{align}
&\EX\left[\sum_{l,j,s:r_{j,s}>0} Q_{l,j,s}(t) \left(\tilde{A}_{l,j}(t) - \hat{\Delta}_{l,j,s}(t)\right)|\mbf{Q}(t)\right] \nonumber \\
&\quad= \EX\left[\sum_{l,j:r_{j,s}>0} \min_l \left(\sum_s Q_{l,j,s}(t)\right) \tilde{A}_{j}(t)  -
\sum_{l,j,s:r_{j,s}>0} Q_{l,j,s}(t) \hat{\Delta}_{l,j,s}(t)|\mbf{Q}(t)\right] \nonumber \\
&\quad\le \EX\left[\sum_{l,j:r_{j,s}>0}  Q_{l,j,s}(t) \frac{\lambda_j^l}{\lambda_j}\tilde{A}_{j}(t)  -
\sum_{j,s:r_{j,s}>0} Q_{l,j,s}(t) \hat{\Delta}_{l,j,s}(t)|\mbf{Q}(t)\right], \nonumber
\end{align}
where $(1-\beta^*) \vlambda = \sum_l (1-\beta^*) \vlambda^l$ for 
$\vlambda^l + \epsilon \mbf{1} \in \C^l$. The remainder of the proof is similar to the approach
for the FD/FND/IND settings, i.e., the MWTA proof for
$(1-\beta^*) \vlambda^l + \epsilon \mbf{1} \in \C^l$ for each $l$.

\end{appendix}

\end{document}